\documentclass{JFM-FLM_Au}

\usepackage{amsmath}
\usepackage{bm}
\usepackage{placeins}

\graphicspath{{figures/}}

\lefttitle{X. Wang, Y. Bao and B. Wang}
\righttitle{Turbulence dynamics in pulsatile pipe flow}

\title{Phase-resolved turbulence dynamics in
cardiac-cycle-inspired pulsatile pipe flow}

\author{Xinyu Wang\aff{1},
  Yan Bao\aff{1,2}
  \and Benlong Wang\aff{1,2}}

\affiliation{\aff{1}School of Ocean and Civil Engineering,
  Shanghai Jiao Tong University, Shanghai 200240, PR China
  \aff{2}Key Laboratory of Hydrodynamics, Ministry of Education,
  Shanghai 200240, PR China}

\corresau{Yan Bao, \email{ybao@sjtu.edu.cn}}

\newcommand{\Reb}{\Rey_b}

\newcommand{\Ub}{U_b}
\newcommand{\utauzero}{u_{\tau,0}}

\newcommand{\phaseavg}[1]{\left\langle #1 \right\rangle}

\newcommand{\tauv}{\tau_{\nu}}
\newcommand{\tauR}{\tau_R}

\newcommand{\tauw}{\tau_w}
\newcommand{\tauL}{\tau_L}
\newcommand{\tauT}{\tau_T}
\newcommand{\tauU}{\tau_U}

\newcommand{\ycross}{y_{\times}^{+}}

\newcommand{\innermode}{\textit{inner mode}}
\newcommand{\outermode}{\textit{outer mode}}
\newcommand{\varicosemode}{\textit{varicose} mode}
\newcommand{\sinuousmode}{\textit{sinuous} mode}
\newcommand{\downtop}{\textit{down-top}}
\newcommand{\topdown}{\textit{top-down}}

\newcommand{\projectfigure}[2][\textwidth]{%
  \IfFileExists{figures/#2.pdf}{%
    \includegraphics[width=#1]{figures/#2.pdf}%
  }{%
    \IfFileExists{figures/#2.png}{%
      \includegraphics[width=#1]{figures/#2.png}%
    }{%
      \fbox{\parbox[c][0.28\textheight][c]{0.9\textwidth}{%
        \centering Figure file pending:\\[0.5em]\texttt{\detokenize{#2.pdf}}}}%
    }%
  }%
}

\makeatletter
\def\ps@headings{%
  \let\@mkboth\markboth
  \def\@oddhead{\hfil{\itshape\@righttitle}\hfil}%
  \def\@evenhead{\hfil{\itshape\@lefttitle}\hfil}%
  \def\@oddfoot{\hfil{\normalfont\cppagefont\thepage}\hfil}%
  \let\@evenfoot\@oddfoot
  \def\sectionmark##1{\markboth{##1}{}}%
  \def\subsectionmark##1{\markright{##1}}%
}
\def\ps@titlepage{%
  \let\@mkboth\@gobbletwo
  \let\@oddhead\@empty
  \let\@evenhead\@empty
  \def\@oddfoot{\hfil{\normalfont\cppagefont\thepage}\hfil}%
  \let\@evenfoot\@oddfoot
  \def\sectionmark##1{}%
  \def\subsectionmark##1{}%
}
\makeatother

\begin{document}

\pagestyle{headings}
\maketitle
\raggedbottom

\begin{abstract}
Direct numerical simulations examine the phase-dependent turbulence dynamics of pipe flow driven by cardiac-cycle-inspired waveforms. Four cases with and without a low-flow plateau span two Reynolds-number ranges. The response follows four stages of turbulent--turbulent reorganisation. Strong acceleration suppresses transverse fluctuations and vortical activity. During acceleration and early deceleration, streamwise-normal-stress production recovers ahead of pressure--strain redistribution and radial momentum transport, allowing energy to accumulate in elongated streaks. In early deceleration, near-wall high-speed streaks develop predominantly varicose disturbances associated with Q4 sweeps, whereas outer-buffer-layer low-speed streaks develop predominantly sinuous disturbances associated with Q2 ejections. The transition exhibits distinct radial sequences: Q2-based local onset occurs progressively later towards the wall, whereas Q4-based onset is nearly simultaneous across the viscous sublayer and buffer layer, with peaks occurring successively outwards. The two responses overlap between the drag minimum and recovery maximum as redistribution intensifies markedly and multiscale momentum transport recovers rapidly. Turbulence then decays and its scales readjust. Where present, the plateau prolongs this decay and weakens the turbulence inherited by the next acceleration stage.  It further delays local instability and lengthens the onset-to-peak intervals of the shear contributions. An unsteady friction decomposition links these phase-dependent dynamics to the drag response, highlighting the role of delayed turbulent shear recovery in sustaining the low-drag interval.
\end{abstract}

\section{Introduction}
\label{sec:introduction}

Turbulent pipe flow is ubiquitous in oil, gas and water-transport systems, where multiscale velocity fluctuations increase wall friction and pumping power. Alongside passive modifications to wall properties, time-dependent wall motion and streamwise forcing offer active means of modifying near-wall turbulence to reduce drag while maintaining transport capacity.

\subsection{Drag reduction by unsteady forcing}
\label{sec:active-control}

Wall oscillation and rotation are established open-loop control strategies, with drag reduction demonstrated in both channel \citep{miyake1997_wall_oscillation} and pipe flows \citep{quadrio2000_pipe_oscillation}. \citet{touber2012_streak_modification} showed that the Stokes layer generated by spanwise wall oscillation perturbs near-wall streaks and their regeneration cycle. Above the viscous sublayer, this layer also creates a ``shielding'' effect that suppresses wall-normal splatting and antisplatting. These changes link active drag reduction to the organisation of energy within streaks and its wall-normal transport.

Periodic modulation of the flow rate or streamwise pressure gradient provides another means of controlling turbulence over the forcing cycle. Although large-amplitude pulsation can reduce the cycle-averaged drag \citep{manna2012_pulsating_part1,mao1994_oscillatory_drag}, the mean velocity, Reynolds stresses and turbulent structures remain strongly phase-dependent \citep{manna2015_pulsating_part2}, with a finite lag between the wall-shear-stress and turbulence responses \citep{liu2024_time_delayed}. Drag reduction does not necessarily imply energy savings.

\begin{figure}
  \centering
  \projectfigure{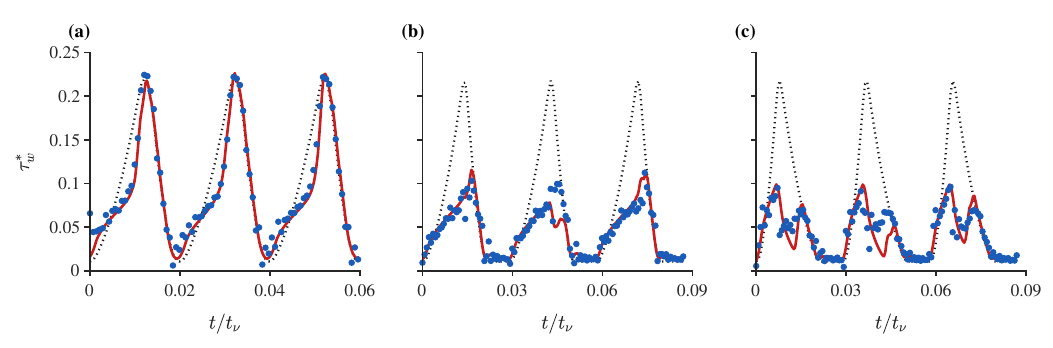}
  \caption{Normalised wall shear stress for the driving waveforms examined by
  \citet{scarselli2023_cardiac_cycle}: experiments (blue circles), DNS
  (red lines) and quasi-steady estimates (black lines), where
  \(\tau_{w,\mathrm{qs}}^*(t)
  =0.079\,\Rey_m(t)^{-0.25}[U_m(t)/U_{\min}]^2\).
  The stress normalisation follows the cited study.
  The panels correspond to different combinations of acceleration and
  minimum-flow plateau durations. Redrawn from data reported by
  \citet{scarselli2023_cardiac_cycle}.}
  \label{fig:literature-waveforms}
\end{figure}

Suitably designed waveforms can reduce both drag and pumping-energy consumption \citep{iwamoto2007_pulsating_channel,foggi_rota2023_on_off,scarselli2023_cardiac_cycle,kobayashi2021_machine_learning,kranz2025_bayesian,rafiei2026_waveform_parameters}.
Motivated by the aortic flow-rate waveform, \citet{scarselli2023_cardiac_cycle} applied a periodic drive comprising rapid acceleration, slower deceleration and a plateau at the minimum flow rate. At a cycle-averaged Reynolds number of \(\Rey=8600\) and \(\mathit{Wo}=14\), their experiments yielded drag reduction and energy savings of \(27\,\%\) and \(9\,\%\), respectively, demonstrating the potential of this waveform at relatively high Reynolds numbers. Figure~\ref{fig:literature-waveforms} summarises the wall-shear-stress histories obtained by varying the acceleration duration \(T_a\) and minimum-flow plateau duration \(T_r\) while holding the maximum and minimum Reynolds numbers and the total acceleration--deceleration duration fixed. The optimum waveform produces a pronounced wall-stress undershoot followed by recovery. Their parametric study indicated that, within the parameter range examined, appreciable energy savings required \(T_a\lesssim0.01D^2/(4\nu)\) and that an optimum plateau duration existed, with \(4\nu T_r/D^2\approx0.01\). Rapid acceleration and the minimum-flow plateau therefore jointly determine how the low-drag state forms and persists across successive cycles.

\citet{kranz2025_bayesian} used Bayesian optimisation to identify energy-efficient bulk-velocity waveforms, considering both cardiac-cycle-inspired piecewise-linear waveforms and truncated Fourier representations. For a representative case at a cycle-averaged Reynolds number of \(\Rey=8600\) and \(\mathit{Wo}=10\), the optimal waveform reduced the pumping-energy consumption by \(22\,\%\) and the drag by \(37\,\%\). Their analysis of intra-cycle dynamics also showed turbulence suppression before acceleration, delayed turbulence recovery, and localisation of turbulent kinetic energy and its production closer to the core.

These studies establish the importance of waveform parameters, the relative timing of low drag and high flow, and turbulent memory \citep{rafiei2026_waveform_parameters,scarselli2023_cardiac_cycle,kranz2025_bayesian}. Understanding the pronounced wall-stress undershoot and subsequent recovery under optimised cardiac-cycle-inspired driving requires examining how turbulence statistics, momentum transport and turbulent structures, particularly streaks, evolve throughout the cycle. Studies of isolated transients provide a starting point for examining these dynamics.

\subsection{Transient acceleration, deceleration and the dynamics of pulsatile pipe flow}
\label{sec:transient-dynamics}

Following rapid acceleration, turbulent flow commonly evolves through three stages \citep{he2011_accelerating_pipe,he2015_transient_transition,he2013_transient_channel}. The pre-existing turbulence first responds weakly and exhibits quasi-laminarisation as near-wall streaks elongate. The elongated streaks then become unstable and generate turbulent spots, which grow and merge to establish fully developed turbulence at the new Reynolds number. This sequence resembles bypass transition induced by free-stream turbulence \citep{he2013_transient_channel}.

Studies of bypass transition in spatially developing boundary layers provide a structural basis for interpreting these streak instabilities. \citet{jacobs2001_bypass_transition} associated transition with the penetration of low-frequency disturbances, streak formation and the subsequent merging of turbulent spots. \citet{hack2014_streak_instabilities} distinguished an \innermode, induced by the local overlap of high- and low-speed streaks, from an \outermode, associated with streamwise meandering near the crest of a lifted low-speed streak. \citet{andersson2001_streak_breakdown} further showed that the \sinuousmode{} and \varicosemode{} of low-speed streaks have different critical conditions. \citet{nolan2010_quadrant_transition} found that Q2 ejections make an important contribution to the Reynolds shear stress before transition and during early breakdown, whereas near-wall Q4 sweeps strengthen with increasing Reynolds number and intermittency. These results concern spatially developing external boundary layers and are used here only as qualitative references for streak instability and the recovery of momentum transport in pulsatile pipe flow.

For the early stage of acceleration, \citet{he2011_accelerating_pipe,he2011_early_stage} developed a quasi-laminar model for predicting the wall shear stress, while \citet{mathur2018_temporal_acceleration} interpreted the pre-acceleration turbulence as the initial perturbation to a newly developing boundary layer. Whether the subsequent stages remain distinct depends on both the final-to-initial Reynolds-number ratio and the acceleration rate \citep{he2015_transient_transition,jung2017_accelerating_channel}. \citet{guerrero2021_accelerating_pipe} used an FIK framework to quantify the components of transient drag. \citet{guerrero2025_spectral_net_force} further linked scale-dependent Reynolds shear transport and the net turbulent force to the friction response during isolated pipe-flow acceleration. \citet{sundstrom2018_pulsating_similarity} showed that pulsatile and isolated accelerating flows exhibit similar early responses over a finite time interval.

Decelerating flow is likewise strongly non-equilibrium. At the onset of deceleration, the mean velocity profile begins to adjust to the lower velocity near the wall, while the outer region temporarily retains the form of the initial turbulent base flow \citep{lee2018_decelerating_pipe}. The turbulence may first remain approximately frozen before decaying \citep{maruyama1976_transient_pipe,ariyaratne2010_decelerating_pipe}, and the Reynolds normal stresses respond at different rates \citep{seddighi2011_ramp_up_down}. In turbulent channel flow following a sudden reduction in pressure gradient, \citet{chung2005_pressure_gradient} characterised the response in terms of fast and slow relaxation, as reflected in the friction and centreline velocities. \citet{guerrero2023_decelerating_pipe} divided rapidly decelerating pipe flow into inertial, friction-recovery, turbulence-decay and core-relaxation stages and used an unsteady FIK decomposition to identify the evolution of the turbulent and unsteady contributions. A decomposition of turbulent-kinetic-energy production further indicated that turbulence develops differently under acceleration and deceleration \citep{sundstrom2018_laminar_similarities}.

Periodic forcing couples these transient responses: each acceleration, deceleration or plateau stage modifies the initial condition of the next. Transition thresholds and turbulence persistence depend on forcing frequency \citep{xu2017_pulsating_transition}, acceleration and deceleration rates, and the duration of the low-flow interval \citep{moron2022_waveform_transition}. These studies examine laminar base states, transition thresholds and turbulence survival, or the evolution and transport of momentum and energy. Within this body of work, less attention has been given to systematically relating structural evolution to transport dynamics using the turbulent--turbulent transition framework developed for accelerating flows. For initially turbulent channel flow, however, \citet{taylor2024_turbulent_transient} applied this framework to large-amplitude sinusoidal forcing, identifying residual-decay, pre-transition, transition, post-transition and turbulence-decay stages. Their findings provide a reference for the present study. The drag reduction and energy savings demonstrated under cardiac-cycle-inspired pipe forcing \citep{scarselli2023_cardiac_cycle,kranz2025_bayesian} motivate further investigation of the underlying mechanisms.

\subsection{The present study}
\label{sec:objectives}

The present study examines the phase-resolved turbulence dynamics of pipe flow under cardiac-cycle-inspired forcing based on the optimal waveform reported by \citet{scarselli2023_cardiac_cycle}. The analysis spans the full forcing cycle, from turbulence suppression and streamwise-energy accumulation during acceleration, through instability and recovery, to subsequent decay and relaxation. Within this framework, a central question is how Q2 and Q4 contributions to momentum transport become established at different wall distances as streaks become unstable and break down. The low-flow plateau is examined through its effects on the turbulent state inherited by the next acceleration stage, the onset and development of instability, and the persistence of the low-drag window. These analyses relate turbulence evolution within and across successive cycles to the mechanisms underlying drag reduction and energy savings.

Direct numerical simulations are performed for forcing waveforms with and without a low-flow plateau over two Reynolds-number ranges. Instantaneous vortical and streak structures, premultiplied spectra and cospectra, and Q2/Q4 statistics characterise structural evolution, energetic scales and changes in shear transport. Phase-averaged velocity profiles, Reynolds stresses and stress-budget terms quantify the mean-flow response and changes in fluctuation energy and momentum transport throughout the cycle. An unsteady FIK decomposition \citep{fukagata2002_skin_friction,guerrero2021_accelerating_pipe,guerrero2025_spectral_net_force} separates the laminar, turbulent and unsteady contributions to wall friction and provides an integral interpretation of the drag response.
\section{Numerical methodology}
\label{sec:numerical-method}

\subsection{Control waveform and case design}
\label{sec:waveform-cases}

We consider incompressible flow through a circular pipe of diameter \(D\), radius \(R\) and axial length \(L_z\). Cylindrical coordinates \((r,\theta,z)\) denote the radial, azimuthal and streamwise directions, with corresponding velocity components \((u_r,u_\theta,u_z)\). Time is reported in seconds using the common scales \(R=0.5\,\mathrm{m}\) and \(\nu=3.3\times10^{-4}\,\mathrm{m^2\,s^{-1}}\). The viscous time scale is \(t_\nu=R^2/\nu\). The prescribed bulk velocity \(\Ub(t)\) defines the bulk Reynolds number \(\Reb(t)=\Ub(t)D/\nu\). Over each period \(T\), \(\Reb\) increases linearly from 3400 to its maximum and then decreases linearly to 3400. In the plateau cases, it remains at this minimum until the next cycle.
Slope changes are smoothed locally using quadratic segments with a half-width of \(0.025\,\mathrm{s}\), ensuring continuity of \(\Reb(t)\) and its first derivative.

\begin{figure}
  \centering
  \projectfigure{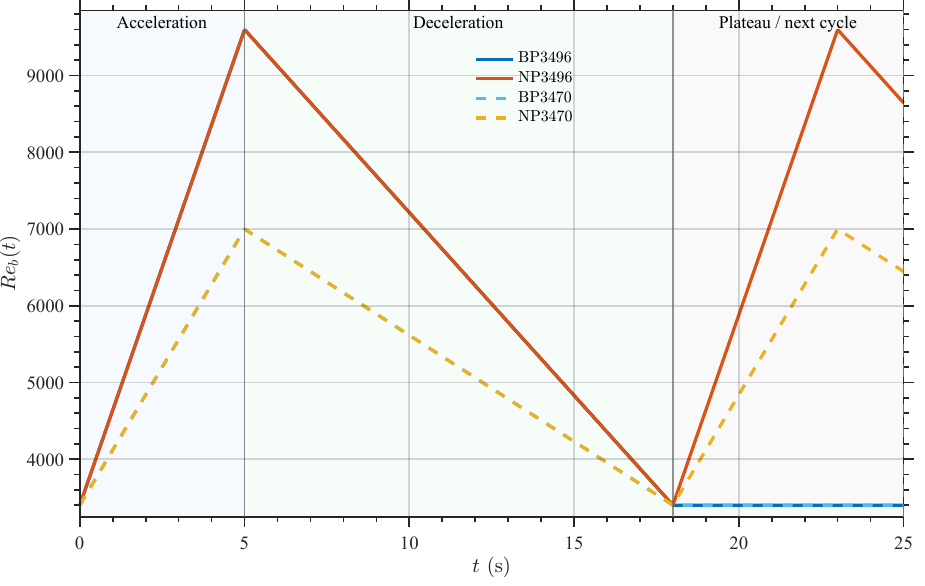}
  \caption{Bulk-Reynolds-number waveforms \(\Reb(t)\) for BP3496, NP3496, BP3470 and NP3470 over \(0\leq t\leq25~\mathrm{s}\). The initial acceleration--deceleration segments coincide within each BP/NP pair; shorter-period cases extend into the next cycle.}
  \label{fig:case-waveforms}
\end{figure}

We use four cases to compare plateau and no-plateau forcing over two Reynolds-number ranges, following \citet{scarselli2023_cardiac_cycle}. BP3496 and NP3496 span \(3400\leq\Reb\leq9600\), while BP3470 and NP3470 span \(3400\leq\Reb\leq7000\). BP (best plateau) denotes forcing with the adopted plateau duration, and NP denotes forcing without a plateau.
In all cases, acceleration and deceleration end at \(5~\mathrm{s}\) (\(0.0066t_\nu\)) and \(18~\mathrm{s}\) (\(0.024t_\nu\)) after the start of each cycle, respectively. The BP cases add a \(7~\mathrm{s}\) (\(0.00924t_\nu\)) plateau, giving a period of \(25~\mathrm{s}\), whereas the NP cases have an \(18~\mathrm{s}\) period. Within each BP/NP pair, the acceleration--deceleration segment and Reynolds-number range are fixed, while adding the plateau changes the period and cycle-averaged flow rate (figure~\ref{fig:case-waveforms}).

\subsection{Definitions of drag reduction and energy saving}
\label{sec:performance-metrics}

Starred shear stresses are defined by \(\tau_i^*=\tau_i/(\rho\utauzero^2)\), where \(\utauzero\) is the steady-reference friction velocity at \(\Reb=3400\). Drag reduction is measured relative to fully developed steady turbulent flow at the same cycle-averaged Reynolds number:
\begin{equation}
  DR=
  \frac{{\tau_{w,\mathrm{steady}}^*}
  -\overline{\tau_w^*}}
  {{\tau_{w,\mathrm{steady}}^*}}.
  \label{eq:drag-reduction}
\end{equation}
Here, \(\overline{\tau_w^*}\) is the cycle-averaged dimensionless wall shear stress of the pulsatile flow. The steady-reference friction coefficient at \(\overline{\Reb}\) follows the Blasius relation \(C_f=0.079\,\overline{\Reb}^{-1/4}\), with \(\tau_{w,\mathrm{steady}}=\rho C_f\overline{\Ub}^{\,2}/2\).

The corresponding net energy savings are defined as
\begin{equation}
  ES=
  \frac{P_{\mathrm{steady}}-\overline{P}}
  {P_{\mathrm{steady}}},
  \label{eq:energy-savings}
\end{equation}
where \(P_{\mathrm{steady}}\) and \(\overline{P}\) are the steady and cycle-averaged pulsatile pumping powers per unit axial length at the same \(\overline{\Reb}\). In the statistically periodic state, the latter equals the cycle-averaged wall-friction power.

\subsection{Statistical decomposition and averaging}
\label{sec:averaging}

An arbitrary flow quantity \(f\) is represented using the triple decomposition
\begin{equation}
  \begin{aligned}
  f(\boldsymbol{x},t)
  &=\frac{1}{N_c}\sum_{k=1}^{N_c}
    f\!\left(\boldsymbol{x},t_\phi+(k-1)T\right)
    +f'(\boldsymbol{x},t)\\
  &=\overline{f}(\boldsymbol{x})
    +\widetilde{f}(\boldsymbol{x},\phi)+f'(\boldsymbol{x},t),
  \end{aligned}
  \label{eq:triple-decomposition}
\end{equation}
where \(t_\phi\) is the phase-\(\phi\) sample time in the first statistical cycle. The prime denotes fluctuations about the pointwise phase mean; the overbar and tilde denote its cycle average and periodic modulation.

Reynolds stresses and transport terms are evaluated pointwise from the accumulated moments and gradients before the resulting fields are averaged in the streamwise direction and subsequently in the azimuthal direction. For \(N_c\) statistical cycles and \(N_{\mathrm{phase}}\) samples per cycle, the reported phase average is
\begin{equation}
  \phaseavg{f(r,t_n)}
  =\frac{1}{2\pi L_z}
  \int_0^{2\pi}\int_0^{L_z}
  \left[\frac{1}{N_c}\sum_{k=1}^{N_c}
    f\!\left(z,r,\theta,t_n+(k-1)T\right)\right]
  \,\mathrm{d}z\,\mathrm{d}\theta,
  \label{eq:phase-average}
\end{equation}
with \(n=1,\ldots,N_{\mathrm{phase}}\). The sampling interval is \(\Delta t=0.5\,\mathrm{s}\), giving \(N_{\mathrm{phase}}=50\) for BP cases and \(36\) for NP cases. Labels run from PHS-0 at the onset of acceleration to PHS-\((N_{\mathrm{phase}}-1)\); PHS-\(N_{\mathrm{phase}}\) denotes the next cycle boundary.

\subsection{Numerical discretisation and computational domain}
\label{sec:numerics}

The simulations use the spectral-element--Fourier DNS solver \textsc{Semtex} \citep{blackburn2004_sem_fourier,blackburn2019_semtex}. Unless stated otherwise, inner variables use the fixed scales \(\utauzero\) and \(\nu\); for example, \(y^+=(R-r)\utauzero/\nu\). Stresses and budget terms are discussed in dimensional form unless a normalisation is explicitly indicated.

\begin{table}
  \centering
  \small
  \setlength{\tabcolsep}{4pt}
  \begin{tabular}{lcccccccc}
    \toprule
    Case & $L_z$ & $N_z$ & $\Delta z_m^+$ & $\Delta r_m^+$
         & $\Delta\theta_m^+$ & $\overline{\Reb}$ & $DR$ (\%) & $ES$ (\%) \\
    \midrule
    BP3496 & $70R$ & 960 & 18.16 & 0.470--7.80 & 2.45--9.42 & 5632 & 14.10 & 4.96 \\
    NP3496 & $34R$ & 480 & 18.39 & 0.489--8.13 & 2.55--9.82 & 6500 & 5.05 & $-1.31$ \\
    BP3470 & $70R$ & 768 & 18.75 & 0.495--7.43 & 2.59--8.95 & 4696 & 9.39 & 5.03 \\
    NP3470 & $34R$ & 480 & 15.14 & 0.403--6.69 & 2.10--8.09 & 5200 & 4.84 & 1.88 \\
    \bottomrule
  \end{tabular}
  \caption{Case parameters, spatial resolution and cycle-averaged performance.
  Here $L_z$ and $N_z$ denote the streamwise domain length and Fourier-mode
  count. Wall-unit spacings use $u_{\tau,m}$, the maximum phase-averaged
  friction velocity; radial and
  azimuthal spacings are given as ranges. The mean Reynolds number $\overline{\Reb}$ is
  calculated from the realised bulk velocity. $DR$ and $ES$ are defined by
  equations~\eqref{eq:drag-reduction} and \eqref{eq:energy-savings}.}
  \label{tab:case-discretisation}
\end{table}

The cross-section is discretised using 256 wall-refined quadrilateral spectral elements with nine Gauss--Lobatto--Legendre points in each local coordinate direction, except for BP3470, which uses eight. A Fourier expansion is used in the periodic streamwise direction; no-slip and impermeability conditions are imposed at the wall. Time integration employs a second-order semi-implicit velocity-correction scheme, with the nonlinear convective terms treated explicitly and the viscous terms implicitly. The maximum CFL number does not exceed unity in any case.
Preliminary computations with \(L_z=10R\), following \citet{scarselli2023_cardiac_cycle} and \citet{kranz2025_bayesian}, constrained the evolution of very long streaks. Domain lengths were therefore selected using premultiplied streamwise-velocity spectra and streamwise two-point correlations.
Table~\ref{tab:case-discretisation} lists the domain lengths and discretisations for all cases. BP3496 and NP3496 have comparable wall-unit grid spacings.

All simulations are initialised from a statistically steady turbulent field at \(\Reb=3400\), and the first five complete cycles are discarded. The flow remains turbulent throughout; here, ``transition'' denotes turbulent--turbulent reorganisation and local streak instability.
The results use 20 statistical cycles for BP3496 and BP3470, and 25 cycles for NP3496 and NP3470. In each case, the phase-averaged wall-shear-stress curves and Reynolds-stress profiles changed little when the final five cycles were excluded from the statistical sample.

Grid sensitivity is assessed for NP3496, which has the largest peak wall shear stress among the four cases. The baseline grid \(G_0\) (\(N_z=480\)) is compared with a streamwise-refined grid \(G_z\) (\(N_z=576\)) and a cross-sectionally refined grid \(G_p\) (\(N_z=480\)), in which the number of Gauss--Lobatto--Legendre points in each local coordinate direction is increased from nine to ten. All three grids are compared using ten-cycle averages.

\begin{figure}
  \centering
  \projectfigure{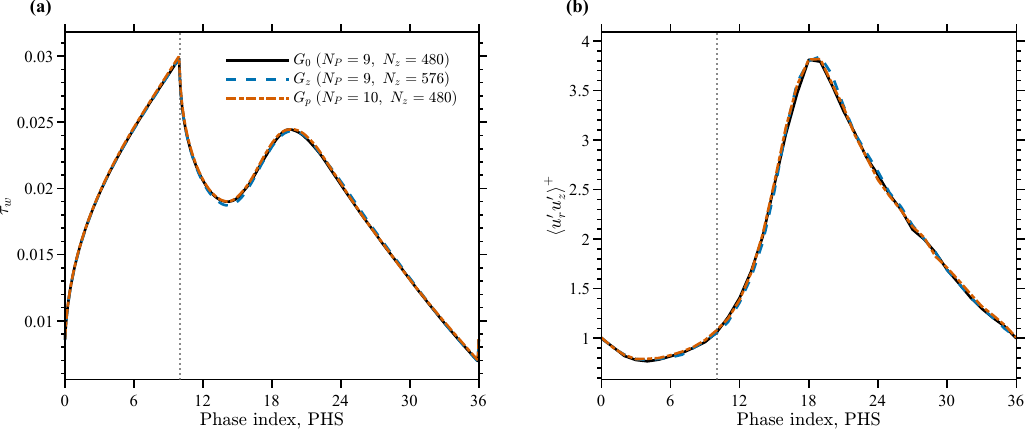}
  \caption{Phase statistics for case NP3496 on grids $G_0$, $G_z$ and $G_p$, each averaged over ten cycles: (a) wall shear stress $\tauw$ and (b) Reynolds shear stress $\phaseavg{u_r'u_z'}^+$ at $y^+=15$. The vertical dashed line marks the end of acceleration at PHS-10.}
  \label{fig:grid-refinement}
\end{figure}

For both refinement paths, the full-cycle relative \(L_2\) differences from \(G_0\) are below \(0.7\,\%\) for the phase-averaged wall shear stress and \(1.9\,\%\) for the Reynolds shear stress at \(y^+=15\) (figure~\ref{fig:grid-refinement}). Peak times differ by no more than one sampling interval. These comparisons indicate low sensitivity of the two NP3496 statistics to the refinement paths tested.

\subsection{Post-processing and verification}
\label{sec:postprocessing}

Post-processing scripts for Reynolds stresses, transport-equation budgets and spectra were developed with assistance from ChatGPT and Codex (OpenAI). For steady turbulent pipe flow at \(\Reb=5300\), the Reynolds-stress profiles and streamwise normal-stress budget agreed well with the reference data of \citet{elkhoury2013_pipe_dns}. The premultiplied streamwise spectra of \(u_z'\) also showed good agreement with the data of \citet{yao2023_pipe_dns}. The spectral implementation was further verified using analytical test signals and discrete Parseval checks.

\section{Drag reduction and energy saving}
\label{sec:results}
\label{sec:drag-energy}

\begin{figure}
  \centering
  \projectfigure{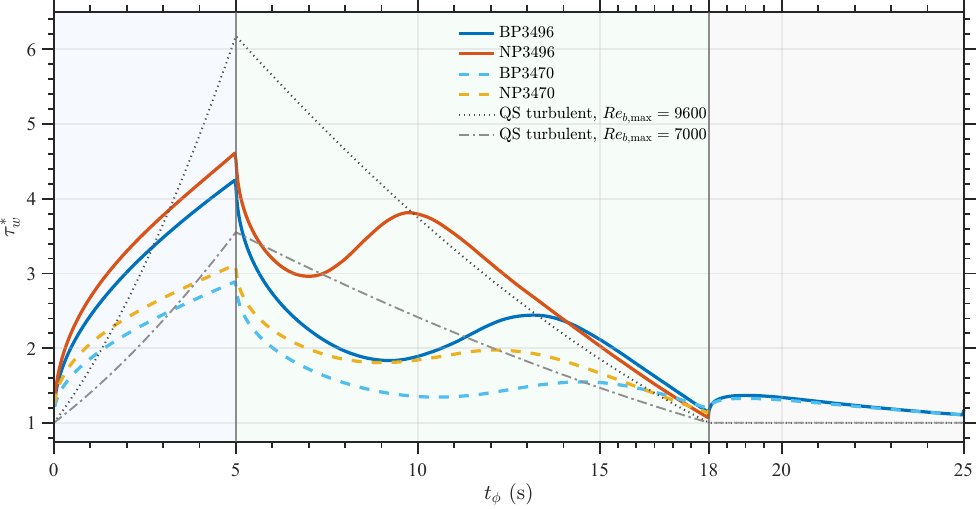}
  \caption{Normalised phase-averaged wall shear stress \(\tauw^*\) as a function of the cycle time
  \(t_\phi\) for the four cases. Solid lines denote cases BP3496 and NP3496
  (\(\Rey_{b,\max}=9600\)), dashed lines denote cases BP3470 and NP3470
  (\(\Rey_{b,\max}=7000\)), and grey lines are the quasi-steady Blasius
  estimates at the instantaneous \(\Reb\). Shading indicates acceleration,
  deceleration and the plateau at the minimum \(\Reb\); vertical lines mark
  the end of acceleration, the end of deceleration and the beginning of the
  following cycle.}
  \label{fig:wall-shear}
\end{figure}

Figure~\ref{fig:wall-shear} shows the normalised phase-averaged wall shear stress \(\tauw^*\) for the four cases, and table~\ref{tab:case-discretisation} summarises their cycle-averaged drag reduction and net energy savings. The timings of the wall-stress extrema reported below are determined from wall-force records sampled more frequently than the flow fields.

All four cases exhibit cycle-averaged drag reduction, whereas net energy savings are obtained in all cases except NP3496. Within each Reynolds-number range, the plateau case yields higher \(DR\) and \(ES\) than its no-plateau counterpart. Pumping-energy savings depend on the relative timing of \(\tauw\) and \(\Ub\), since the cycle-averaged pumping power is proportional to \(\overline{\tauw\Ub}\).
Despite differences in the magnitudes and timings of the extrema, the wall-stress response follows four stages in all cases: growth during acceleration (stage~I), an undershoot to a local minimum during early deceleration (stage~II), recovery to a local maximum (stage~III), and turbulence decay and relaxation from the recovery maximum to the onset of the next acceleration stage (stage~IV). In the plateau cases, the additional interval at the minimum Reynolds number extends the low-drag window and increases the cycle-integrated energetic benefit. This sequence is consistent with that reported by \citet{scarselli2023_cardiac_cycle}. As illustrated in figure~\ref{fig:literature-waveforms}, the pronounced undershoot is associated with rapid acceleration; lengthening the acceleration interval can weaken or eliminate the local minimum. The present four-stage classification is therefore specific to the forcing time scales considered here.

After an initial adjustment during acceleration, \(\tauw\) rises to a peak near the end of the interval. The acceleration-stage peak remains below the quasi-steady turbulent prediction at the same instantaneous Reynolds number. The deficit is larger in BP3496 and NP3496 than in BP3470 and NP3470; within each Reynolds-number range, the peak is higher in the no-plateau case. For a waveform spanning approximately \(3200\lesssim\Rey\lesssim18000\), \citet{scarselli2023_cardiac_cycle} measured a drag peak of about \(40\,\%\) of the quasi-steady value, compared with approximately \(69\,\%\) and \(82\,\%\) in BP3496 and BP3470. This comparison is consistent with a larger maximum Reynolds number increasing the departure from quasi-steady behaviour and suggests that a larger departure may favour drag reduction.

Once deceleration begins, \(\tauw\) falls to a local minimum and then recovers. BP3496 reaches its minimum at \(t_\phi=9.20~\mathrm{s}\) and its recovery maximum at \(13.18~\mathrm{s}\), compared with \(7.00~\mathrm{s}\) and \(9.75~\mathrm{s}\) in NP3496. 
Within the common acceleration--deceleration segment, each plateau case exhibits a deeper undershoot and longer drag-decrease and recovery stages than its no-plateau counterpart, extending the low-drag window. In every case, the recovery maximum occurs shortly before the second intersection of the measured and quasi-steady curves. This correspondence is empirical. Transition and the drag extrema occur later in BP3470 and NP3470, which also have lower acceleration and deceleration rates \citep{moron2022_waveform_transition}. The present design therefore cannot separate the effects of the maximum Reynolds number from those of the acceleration and deceleration rates.

After the recovery maximum, the drag decays and passes above the quasi-steady curve. Cases with the same maximum Reynolds number approach one another in total drag towards the end of deceleration, although their FIK contributions can remain distinct (\S\ref{sec:fik}). During the plateau, BP3496 and BP3470 show a brief rise followed by a slow decline, with similar qualitative behaviour but different amplitudes and relaxation rates.

The four-stage wall-stress response provides a common temporal framework for the analyses below. Detailed comparisons focus primarily on BP3496 and NP3496 to relate turbulence evolution and momentum transport to the drag response and to examine the role of the low-flow plateau.

In both cases, PHS-10 marks the end of acceleration and the onset of deceleration. The representative flow-field phases near the drag minimum and recovery maximum are PHS-18 and PHS-26 for BP3496, and PHS-14 and PHS-20 for NP3496. Deceleration ends at PHS-36 in both cases. NP3496 immediately enters the next cycle, whereas BP3496 remains on the low-flow plateau until PHS-50. Intermediate phases are included where needed to illustrate evolution within each stage.

\section{Kinematics of turbulence and transition}
\label{sec:kinematics}

\subsection{Evolution of vortical and streak structures}
\label{sec:structures}

\begin{figure}
  \centering
  \projectfigure{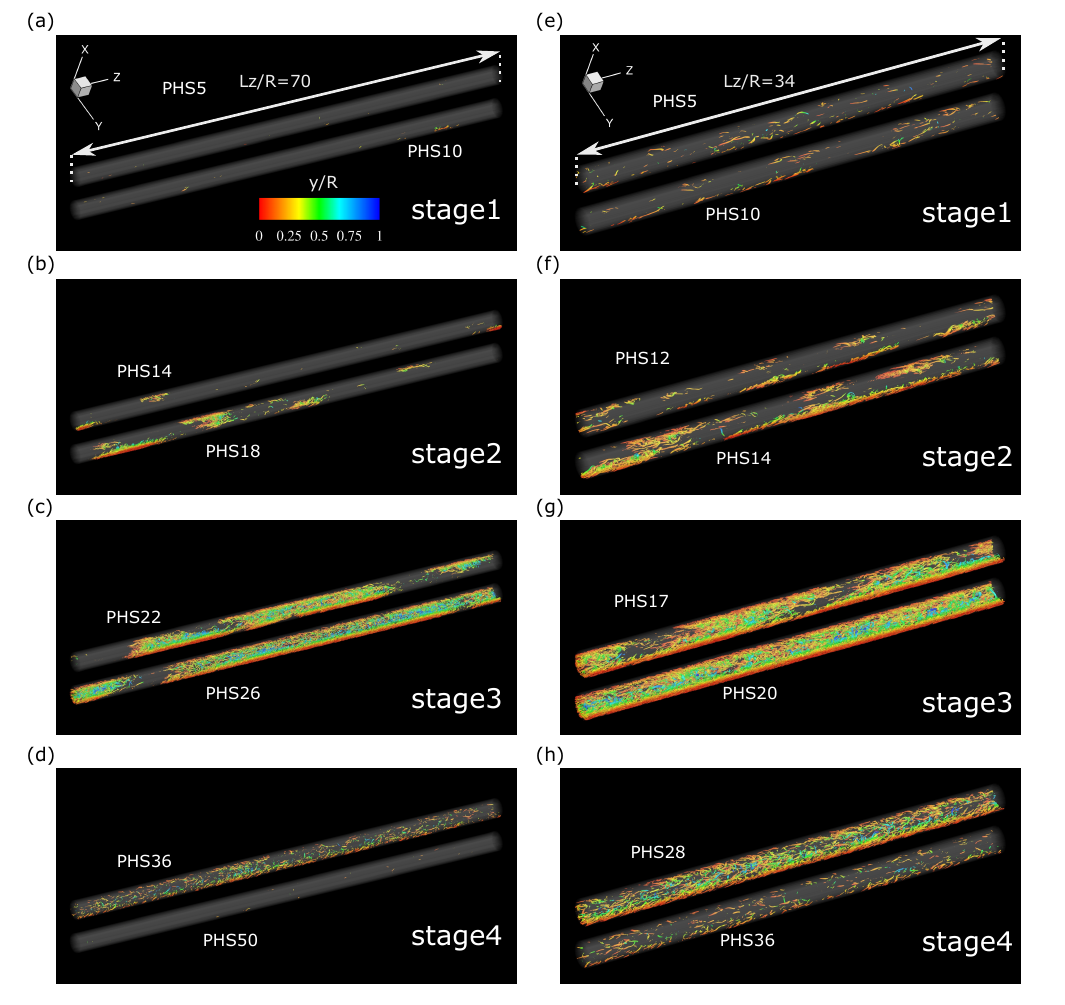}
  \caption{Instantaneous vortical structures in BP3496
  (left column, a--d; $L_z/R=70$) and NP3496
  (right column, e--h; $L_z/R=34$), visualised using
  $\lambda_2^+=-0.125$ isosurfaces over half of the circumference
  and coloured by $y/R$ with a common scale.
  Panels (a,e), (b,f), (c,g) and (d,h) illustrate stages~I--IV,
  respectively. Phase labels are given in the panels.}
  \label{fig:vortical-structures}
\end{figure}

Figure~\ref{fig:vortical-structures} compares instantaneous vortical structures in BP3496 and NP3496 at representative phases illustrating the four stages defined in \S\ref{sec:drag-energy}. 
Corresponding panels are aligned by physical stage rather than by phase number.

Vortical structures are identified using the $\lambda_2$ criterion of \citet{jeong1995_vortex}, with the fixed threshold $\lambda_2^+=-0.125$, where $\lambda_2^+=\lambda_2\nu^2/\utauzero^4$. The complete streamwise domains are plotted at the same figure width; their apparent lengths therefore do not represent comparable axial scales. Supplementary movie~1 shows the evolution of the instantaneous vortical structures in both cases.

Figure~\ref{fig:vortical-structures}(a,e) shows the evolution of vortical structures during acceleration in stage~I. At PHS-5, BP3496 contains markedly fewer identifiable vortical structures than NP3496, indicating that the low-Reynolds-number plateau reduces vortical activity at the onset of the next acceleration stage. By PHS-10, the remaining structures are concentrated near the wall and continue to elongate and align streamwise, while their overall number and density decrease. This behaviour agrees with observations by \citet{guerrero2021_accelerating_pipe} in accelerating pipe flow and with the statistical suppression of the near-wall self-sustaining process.

\begin{figure}
  \centering
  \projectfigure{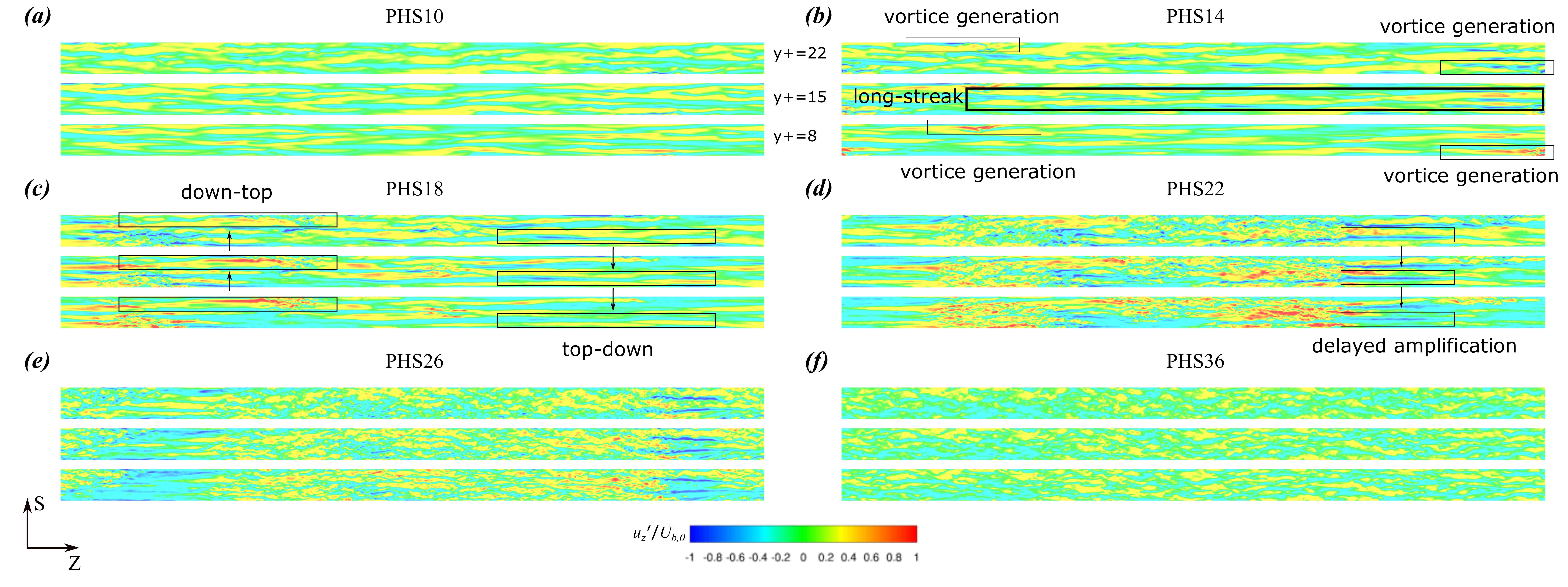}
  \caption{Instantaneous streamwise-velocity fluctuations
$u_z'/U_{b,0}$ in BP3496 at the phases labelled in the panels.
From top to bottom, each panel shows $y^+=22$, 15 and 8.
Here $U_{b,0}=1.122~\mathrm{m\,s^{-1}}$ is the bulk velocity at $\Reb=3400$, and all panels use the common colour
range $[-1,1]$.}
  \label{fig:streak-structures}
\end{figure}

In figure~\ref{fig:vortical-structures}(b,f), vortical structures recover gradually after deceleration begins in stage~II. Recovery remains limited in BP3496 at PHS-14 and becomes pronounced only after local streak instability develops; NP3496 follows the same sequence but responds earlier. New vortical structures appear both very close to the wall (dark red) and in the outer buffer layer, i.e. the outer part of the buffer layer at $y^+\approx20$--30 (yellow). The streak visualisations at nearby times (figure~\ref{fig:streak-structures}(b,c)) show early varicose instability and local breakdown of near-wall high-speed streaks, and sinuous instability and local breakdown of low-speed streaks farther from the wall. Here, varicose and sinuous describe the observed patterns of streak deformation. This broad spatial correspondence links vortex recovery to the instability of both streak types at different wall-normal locations. The vortices are then advected downstream and spread gradually away from the wall. Near their respective drag minima, NP3496 exhibits a broader region of vortex recovery than BP3496.

Figure~\ref{fig:vortical-structures}(c,g) shows a rapid increase in the number and spatial coverage of vortical structures during stage~III, as they spread from the wall towards the pipe centre. The intermediate snapshots at PHS-22 in BP3496 and PHS-17 in NP3496 illustrate this recovery. Some structures have a hairpin- or horseshoe-like appearance and occur primarily where high-speed streak breakdown is concentrated, consistent with hairpin-vortex formation during varicose breakdown reported previously \citep{hack2014_streak_instabilities}.

The fields in figure~\ref{fig:vortical-structures}(d,h) show the overall decay of vortical activity in stage~IV, although local generation persists. Regions sparsely populated by vortices near the recovery maximum also contain relatively continuous low-speed streaks; new vortices appear there when these streaks undergo delayed instability and breakdown. This local production does not alter the overall decay, which begins earlier near the wall than in the outer region. 
At the end of deceleration, PHS-36 in both cases, substantially fewer vortical structures remain than near the recovery maximum. In BP3496, decay continues through the low-flow plateau, and the vortical field has almost vanished by PHS-50, equivalent to PHS-0 of the following cycle.

To connect vortex recovery with streak evolution, figure~\ref{fig:streak-structures} shows instantaneous streamwise-velocity streaks in BP3496 at six representative phases and three wall-normal locations over the full $70R$ domain. Supplementary movie~2 shows the corresponding streak evolution in both BP3496 and NP3496. Here, streak intensity denotes $|u_z'|$, with $u_z'>0$ and $u_z'<0$ identifying high- and low-speed streaks, respectively.

Instantaneous fields extracted at corresponding phases from three cycles, together with auxiliary checks in NP3496, show repeatable instability morphology and similar stage-wise evolution, although the onset phases and quantitative timing differ. NP3496 enters the next acceleration stage immediately after deceleration, without the additional streak decay and scale readjustment observed during the BP3496 plateau. Pronounced streak amplification and local instability occur earlier after the onset of deceleration in NP3496 than in BP3496, reflecting their different initial turbulence states and response times.

During stage I, the streaks elongate and progressively connect along the streamwise direction, while their fluctuation amplitude increases slowly from PHS-0 to PHS-10 (figure~\ref{fig:streak-structures}(a)). The relative increase varies with wall distance: high-speed streaks grow more strongly at $y^+=8$, the two streak types initially show comparable increases in amplitude at $y^+=15$, and low-speed streaks grow more strongly at $y^+=22$. Neither \innermode\ nor \outermode\ instability is identifiable during stage I in either case.

Figure~\ref{fig:streak-structures}(b,c) illustrates streak evolution during stage II in BP3496. The streaks continue to elongate, connect and intensify, with some spanning at least half of the computational domain. As deceleration proceeds, varicose instability and local breakdown first appear in high-speed streaks at $y^+=8$, whereas sinuous instability and local breakdown first appear in low-speed streaks at $y^+=22$. Local breakdown of the two streak types is often spatially associated, with disrupted segments appearing in neighbouring parts of the same disturbed regions. At $y^+=15$, the two streak types appear to become unstable at similar phases. By PHS-18 (figure~\ref{fig:streak-structures}(c)), some high-speed streaks at $y^+=8$ have already broken down, while their breakdown remains less advanced at $y^+=22$. Low-speed streaks show no clear instability at $y^+=8$, pronounced meandering at $y^+=15$ and local breakdown at $y^+=22$. The corresponding local instability points $\phi_{i,\mathrm{onset}}(y)$ are quantified in \S\ref{sec:quadrant}, where their relation to the Q2/Q4 shear-contribution sequences is examined.

The near-wall breakdown of high-speed streaks is consistent with observations in accelerating pipe flow and boundary-layer transition \citep{guerrero2021_accelerating_pipe,mao2017_compressor_transition}. Farther from the wall, the sinuous deformation of low-speed streaks resembles that reported for lifted streaks \citep{jacobs2001_bypass_transition,hack2014_streak_instabilities}. \citet{taylor2024_turbulent_transient} reported vortex clusters forming predominantly around low-speed streaks in sinusoidally forced channel flow. The present comparison across wall-normal locations also identifies near-wall vortex recovery associated with high-speed-streak instability. In the \topdown\ scenario described by \citet{jacobs2001_bypass_transition}, instability of outer low-speed streaks is followed by wallward growth of turbulent spots. \citet{nolan2010_quadrant_transition} observed a wallward shift of the off-wall maximum in turbulent kinetic energy production. For the \innermode, local overlap between high- and low-speed streaks can create strong shear layers or inflectional velocity profiles that favour high-speed-streak deformation and breakdown \citep{hack2014_streak_instabilities}. Differences in the reported instability forms may therefore partly reflect the wall-normal locations examined. These comparisons guide the interpretation of the two breakdown paths.

During rapid transition and drag recovery in stage III, the local instabilities initiated in stage II develop further, and both streak types undergo more widespread breakdown (figure~\ref{fig:streak-structures}(d,e)). Near-wall low-speed streaks undergo delayed amplification and breakdown. High-speed streaks in the outer buffer layer can also break down later, but without a similarly pronounced increase in amplitude. At $y^+=8$, low-speed streaks may exhibit both varicose and sinuous instabilities associated with the inner and outer modes, respectively. Varicose instability also occurs in a few low-speed streaks in the outer buffer layer. Neither instability form is therefore confined to a single wall-normal location, although the principal onset regions of varicose and sinuous instability remain near the wall and in the outer buffer layer, respectively.

Stage IV is dominated by streak decay and scale readjustment, although a few streaks continue to undergo instability and breakdown near the recovery maximum (PHS-26). The streak field subsequently weakens, and its characteristic streamwise scale increases again (figure~\ref{fig:streak-structures}(e,f)). In the plateau cases, this adjustment continues at the minimum Reynolds number, and the characteristic wavelength gradually approaches the steady $\Reb=3400$ value, $\lambda_z^+\approx1000$.

The instantaneous fields thus identify the principal regions and streak-breakdown patterns associated with the inner and outer modes. We next use energy spectra and cospectra to distinguish scale redistribution from wall-normal momentum transport.

\subsection{Premultiplied spectra and cospectra}
\label{sec:spectra}

\begin{figure}
  \centering
  \projectfigure{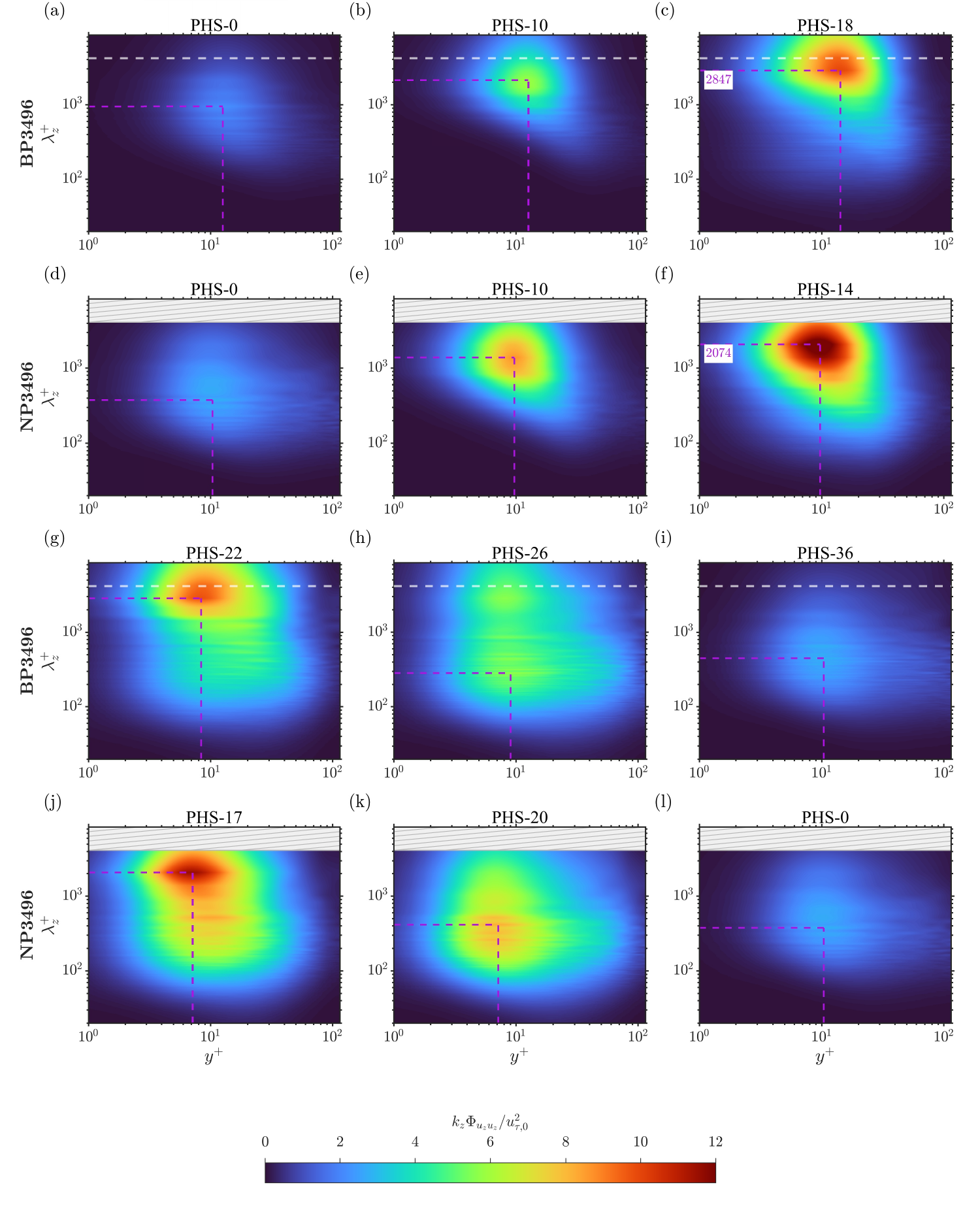}
  \caption{Premultiplied streamwise-velocity spectra $k_z\Phi_{u_z u_z}$ for cases BP3496 and NP3496. Panels (a,d), (b,e), (c,f) and (h,k) represent the onset of stages~I--IV, respectively; panels (g,j) and (i,l) show an intermediate phase in stage~III and the end of deceleration, respectively. Purple dashed lines locate the spectral maxima. The white horizontal dashed line in BP3496 and the hatched region at the top of NP3496 indicate, respectively, the maximum wavelength resolved by the NP3496 domain and wavelengths not covered by that domain.}
  \label{fig:energy-spectra}
\end{figure}

 We examine the phase-averaged premultiplied streamwise-velocity spectrum $k_z\Phi_{u_z u_z}$ and shear-stress cospectrum $k_z\Phi_{u_r u_z}$, where $y$ is the wall-normal coordinate and $\lambda_z$ the streamwise wavelength. The vortex and streak visualisations are drawn from a single representative cycle, whereas the spectra and cospectra are phase averaged; the two forms of evidence are therefore compared only in terms of phase, wall distance and scale. Figure~\ref{fig:energy-spectra} compares the streamwise-velocity spectra of BP3496 and NP3496 at representative phases throughout the cycle. Supplementary movie~3 shows the phase evolution of both spectral quantities together with the wall-shear-stress response.

During stage I, the spectral peak and the principal energetic range shift towards longer wavelengths; over the first few phases, the peak also moves wallwards. The very-long-wavelength peak strengthens continuously, whereas intermediate- and small-scale energy decreases.

During stage II, the spectral peak continues to strengthen and shifts mainly towards longer wavelengths. The peak shifts away from the wall within the buffer layer (from $y^+\approx12.6$ to 14.2 in BP3496). Intermediate- and small-scale energy recovers concurrently. Many very long streaks have not yet fully broken down, so long-scale energy coexists with newly generated intermediate- and small-scale energy, reflecting the concurrent elongation and instability-induced breakdown of streaks.

Near the drag minimum, the spectral-peak wavelengths are approximately $\lambda_z^+=2847$ in BP3496 and 2074 in NP3496. NP3496 also exhibits a stronger shear cospectrum and higher drag. Its shorter local onset-to-peak intervals (\S\ref{sec:quadrant}) support the interpretation that scale redistribution and momentum-transport recovery are concentrated within a shorter time interval during its principal response stage.

Figure~\ref{fig:energy-spectra}(c,f,g,j,h,k) illustrates the spectral evolution during stage~III. The spectral peak weakens slowly while the distribution broadens in wavelength and wall-normal extent. Energy continues to increase near the wall and in the outer region, alongside growing intermediate- and small-scale contributions.
In BP3496, the peak turns wallwards just after the drag minimum and subsequently shifts from $y^+\approx14$ to 8. This shift results from the decay of energy at the original peak location and the delayed amplification of near-wall low-speed streaks. It does not represent wallward migration of an individual streak. Scale redistribution near the wall and in the outer buffer layer corresponds to the two streak-breakdown processes described in \S\ref{sec:structures}; their wall-normal establishment sequence is further constrained by the shear contributions in \S\ref{sec:quadrant}. Beyond the buffer layer, streamwise energy increases simultaneously over a broad range of scales, rather than increasing at long scales while decreasing at intermediate and small scales. The fixed-location spectra below help clarify the structural origin of this outer-region growth.

Figure~\ref{fig:energy-spectra}(h,k,i,l) illustrates the spectral evolution during stage~IV, which is dominated by energy decay and scale readjustment. The spectral peak moves gradually outwards, and the characteristic streamwise scale increases again. By the end of deceleration, the spectral distributions in BP3496 and NP3496 are already similar. 
During the plateau portion of stage~IV in BP3496, the spectrum continues to shift away from the wall and towards the characteristic scale of steady flow at $\Reb=3400$, while its principal distribution narrows in wavelength and wall-normal extent. At the onset of the next acceleration, the peak lies at $y^+\approx12.6$ in BP3496, compared with $y^+\approx10.4$ in NP3496 (figure~\ref{fig:energy-spectra}(a,d)).

\begin{figure}
  \centering
  \projectfigure{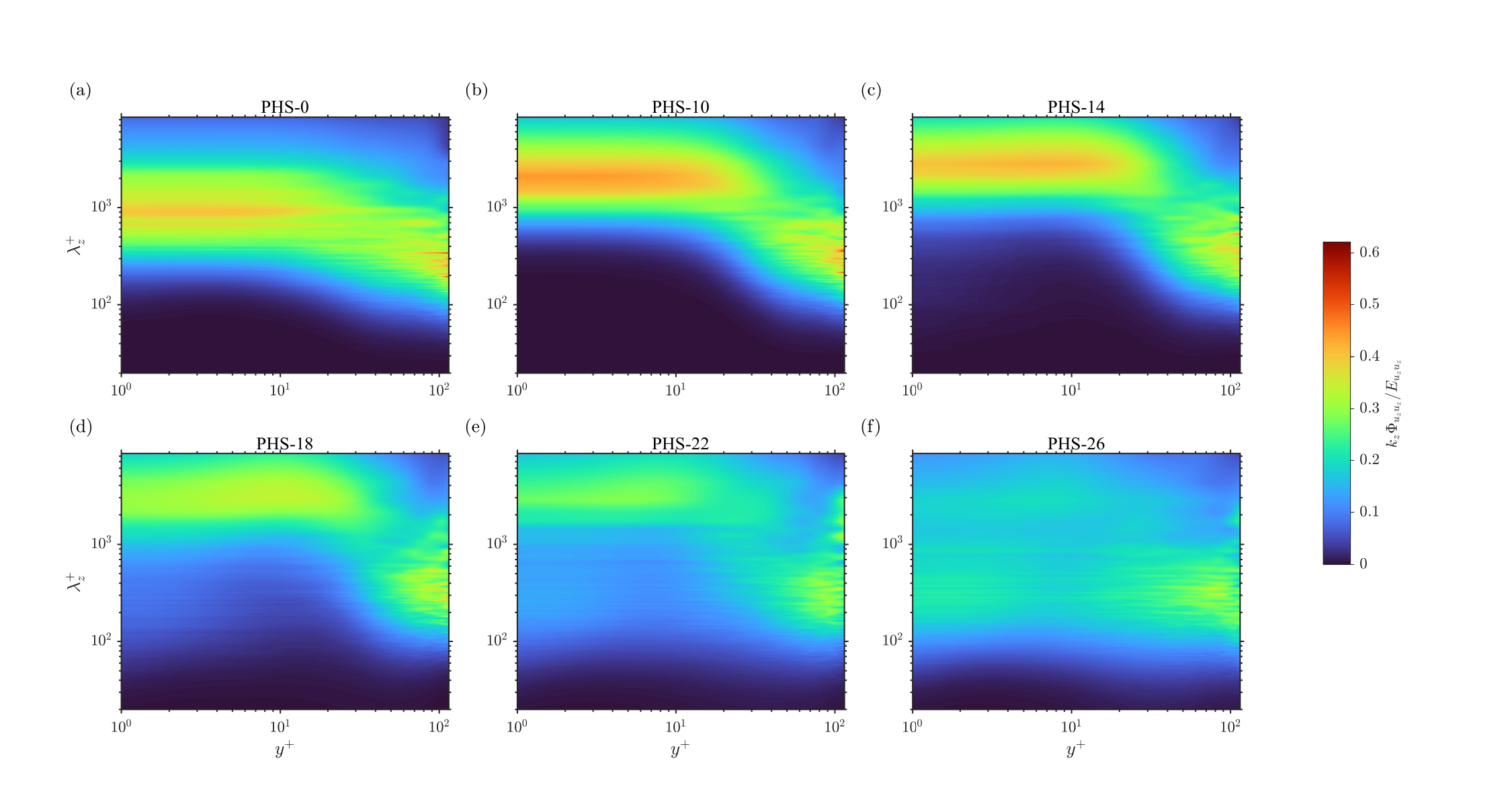}
  \caption{Locally normalised premultiplied spectra $k_z\Phi_{u_z u_z}/E_{u_z u_z}$ in case BP3496 at (a) PHS-0, (b) PHS-10, (c) PHS-14, (d) PHS-18, (e) PHS-22 and (f) PHS-26. Here $E_{u_z u_z}(y,\phi)=\phaseavg{u_z'u_z'}$ denotes the total streamwise fluctuation energy at the same wall distance and phase.}
  \label{fig:normalised-spectra}
\end{figure}

Figure~\ref{fig:normalised-spectra} isolates local scale redistribution by normalising each spectrum by the total streamwise fluctuation energy at the same wall distance and phase. This reveals changes in low-energy regions that are less visible under the common colour scale of figure~\ref{fig:energy-spectra}. From PHS-0 to PHS-10 in BP3496, the spectra shift approximately synchronously towards longer wavelengths throughout $y^+<30$, reflecting streak elongation at each location; changes beyond the buffer layer are weaker. The corresponding spectra for NP3496 (not shown) indicate that elongation is largely confined to $y^+<22$, a narrower wall-normal range than in BP3496.

After deceleration begins, the relative contributions of intermediate and small scales increase at different times at $y^+=8$, 14 and 22 (figure~\ref{fig:normalised-spectra}(d,e)). The near-wall and outer-buffer-layer responses correspond to the two forms of streak instability identified in \S\ref{sec:structures} and accompany increases in the Q4 and Q2 shear contributions, respectively.

By PHS-22, long-scale energy has increased further in the outer buffer and outer regions. Taken together, the fixed-location spectra, the broadband outer-region energy growth in figure~\ref{fig:energy-spectra} and the instantaneous streaks link this increase primarily to the lifting and outward expansion of elongated outer-buffer-layer low-speed streaks. It is not primarily attributable to small outer structures first stretching into new long structures and then breaking down. This scale redistribution is consistent with the distinct Reynolds-stress peak trajectories discussed in \S\ref{sec:stress-evolution}.

\begin{figure}
  \centering
  \projectfigure{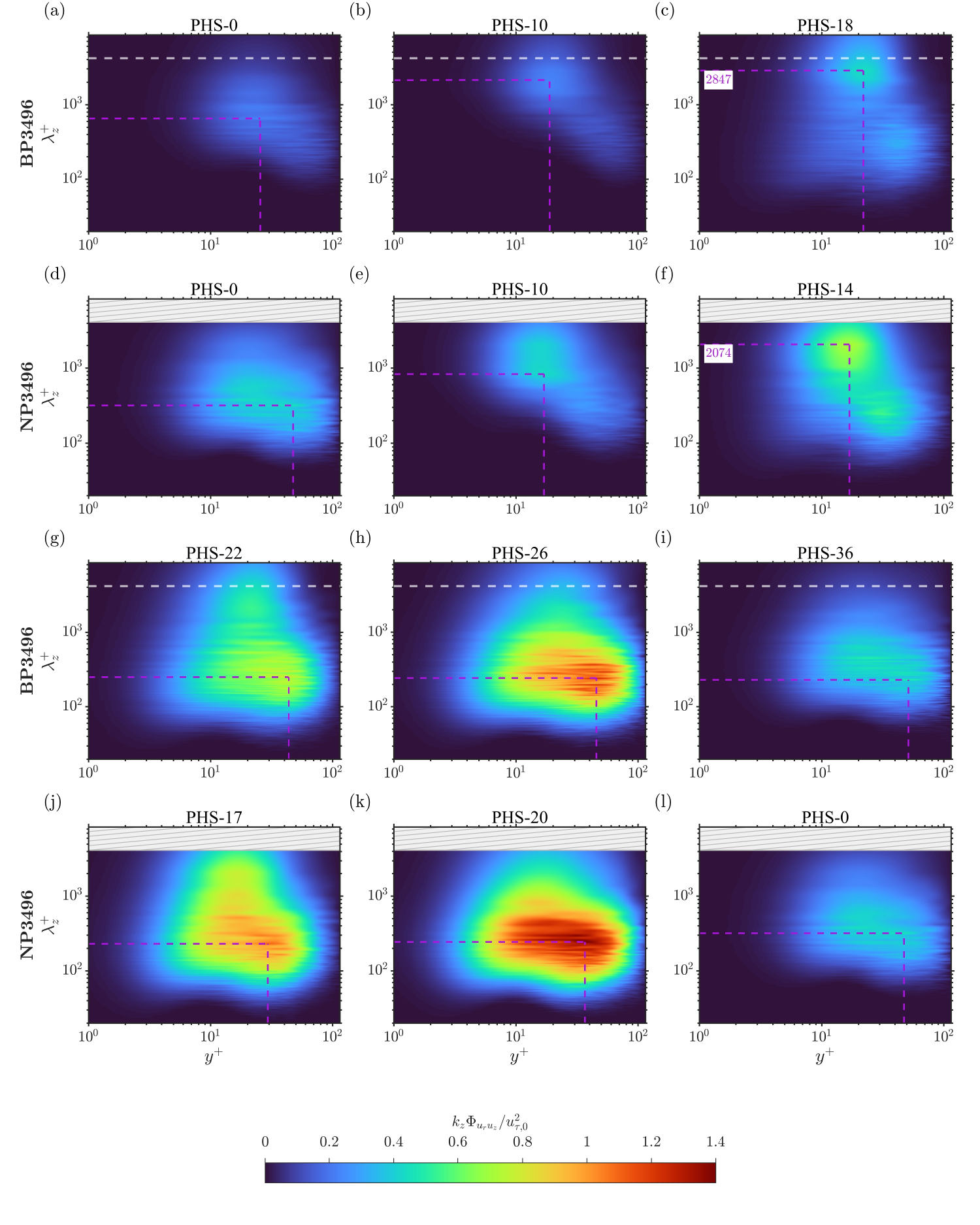}
  \caption{Premultiplied Reynolds-shear-stress cospectra for BP3496 and NP3496. Only positive contributions are shown. Phase pairs, panel arrangement, coordinates and auxiliary markings follow figure~\ref{fig:energy-spectra}. Purple dashed lines mark the maxima, and peak wavelengths are labelled at the drag minimum.}
  \label{fig:shear-cospectra}
\end{figure}

Figure~\ref{fig:shear-cospectra} shows the positive part of the premultiplied shear cospectrum at the same six phase pairs as figure~\ref{fig:energy-spectra}. The small negative contributions are omitted; the plotted quantity is therefore not the signed net cospectrum. The dominant contributions to the energy spectrum and shear cospectrum generally occur at similar wavelengths but at different wall-normal locations.

During stage I, the intermediate- and small-scale contributions to the shear cospectrum weaken markedly, particularly within the viscous sublayer and buffer layer. Although streak elongation shifts its peak towards longer wavelengths, the peak magnitude remains close to its initial value at the end of acceleration. Concentrating streamwise energy in very long streaks therefore does not produce a comparable increase in radial momentum transport; elongated streaks that have not yet become unstable account for the limited increase in long-wavelength shear transport.

In figure~\ref{fig:shear-cospectra}(b,c,e,f), the cospectral peak during stage~II broadly follows the energy-spectrum peak in wavelength but lies farther from the wall. Very long streaks that have not fully broken down sustain and strengthen the long-wave contribution, while intermediate- and small-scale contributions recover within $y^+<30$. These changes provide scale-based evidence for the near-wall and outer-buffer-layer paths.

The shear cospectrum in figure~\ref{fig:shear-cospectra}(c,f,g,j,h,k) strengthens rapidly during stage~III as streak instability and breakdown enhance intermediate- and small-scale contributions, which extend into $y^+<10$. Very long streaks that have not yet become unstable or have only partly broken down sustain a discernible long-wavelength contribution to shear transport. The near-wall and outer-buffer-layer responses overlap during the middle and late stages of deceleration and jointly restore radial momentum transport. The widespread intermediate- and small-scale contributions in the outer region accompany the outward displacement of the Reynolds-shear-stress peak. In both BP3496 and NP3496, the shear-cospectrum peak is strongest near the drag-recovery maximum, by which time intermediate and small scales dominate.

During stage IV, intermediate- and small-scale shear contributions decay progressively. Contributions at $\lambda_z^+<100$ continue to weaken after the drag-recovery maximum, while the cospectral peak weakens and shifts towards longer wavelengths. By the end of deceleration, the two cases show similar peak wavelengths, wall-normal locations and dominant spectral regions in both the energy spectrum and shear cospectrum.
During the subsequent plateau in BP3496, shear contributions within $y^+<30$ and $\lambda_z^+<200$ weaken further and approach zero, while the cospectral intensity decreases at all scales. At the onset of the next acceleration, NP3496 retains a stronger near-wall small-scale contribution, and its cospectrum peaks at $\lambda_z^+\approx320$ in and beyond the outer buffer layer, whereas the peak wavelength in BP3496 has recovered to approximately 660. The difference in scale distribution therefore persists into the next cycle.

\subsection{Q2/Q4 contributions and transition}
\label{sec:quadrant}

\begin{figure}
  \centering
  \projectfigure{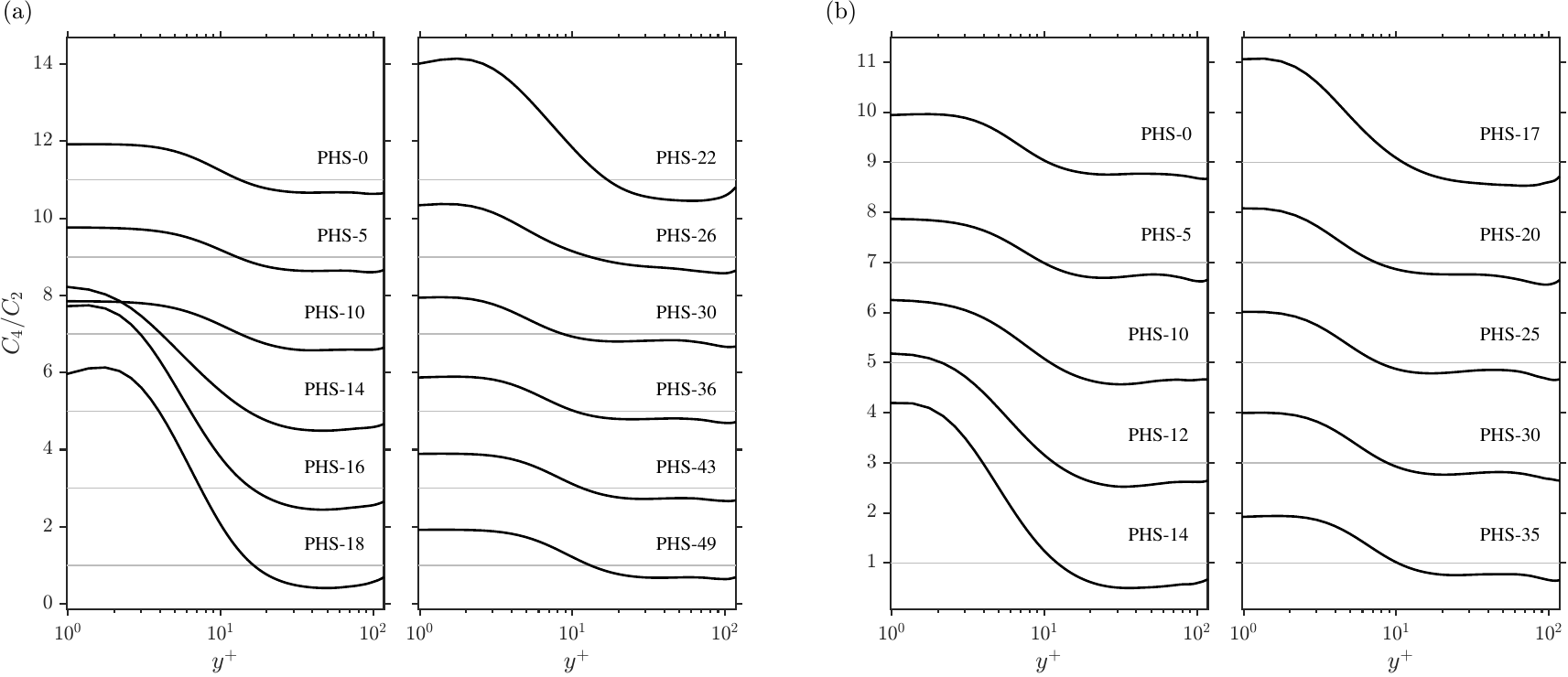}
  \caption{Radial evolution of $C_4/C_2$ over one cycle for (a) case BP3496 and (b) case NP3496. The adjacent axes for each case form a continuous phase sequence, read down the left axis and then down the right axis. Curves are shifted vertically at uniform intervals; each grey horizontal line denotes the correspondingly shifted value $C_4/C_2=1$, and its intersection with a curve gives the crossover location $\ycross$.}
  \label{fig:q4-q2-ratio}
\end{figure}

\begin{figure}
  \centering
  \projectfigure{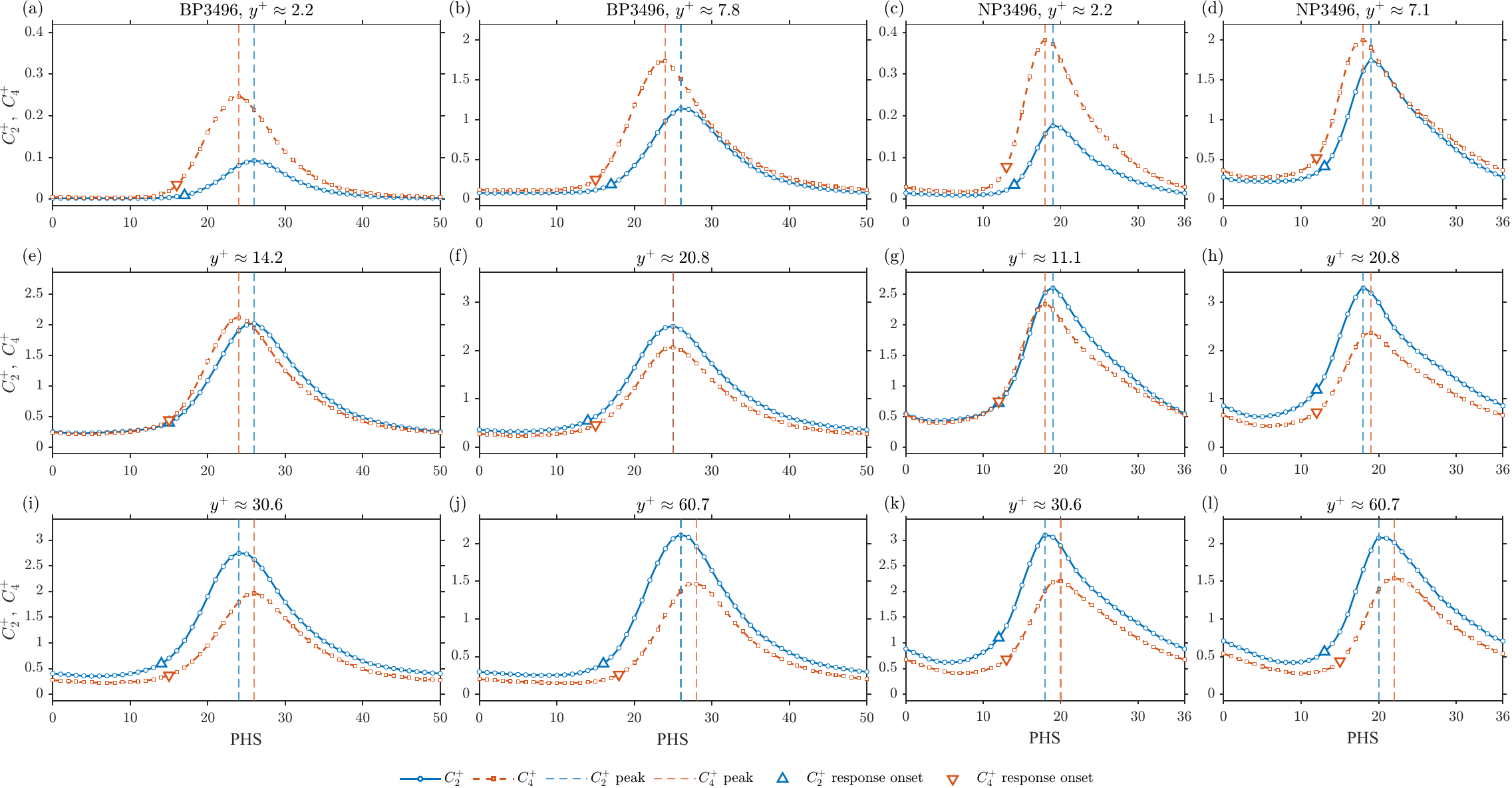}
  \caption{Phase evolution of the Q2 and Q4 contributions, $C_2^+$ and $C_4^+$, at six wall distances in BP3496 and NP3496. The two left and two right panels in each row show BP3496 and NP3496, respectively. The case-specific locations in (b,e) and (d,g) lie near the wall-normal positions of the streamwise-energy spectral peak at the drag extrema. Blue solid lines with circles and orange dashed lines with squares denote $C_2^+$ and $C_4^+$, respectively. Vertical dashed lines of the corresponding colours mark the discrete peak phases; open upward and downward triangles mark the local instability points $\phi_{2,\mathrm{onset}}(y)$ and $\phi_{4,\mathrm{onset}}(y)$, respectively.}
  \label{fig:q2-q4-contributions}
\end{figure}

We next examine the Q2-ejection and Q4-sweep contributions and relate their radial establishment to the scale redistribution identified by the phase-averaged spectra and shear cospectra. Instantaneous streak and vortex fields provide qualitative support only. Figure~\ref{fig:q4-q2-ratio} shows the radial distribution of $C_4/C_2$, while figure~\ref{fig:q2-q4-contributions} compares the phase evolution of $C_2$ and $C_4$ at six wall distances, together with their local instability points $\phi_{i,\mathrm{onset}}(y)$ and discrete peak phases.

The contributions of Q2 ejections and Q4 sweeps to the phase-averaged Reynolds shear stress are defined as
\begin{equation}
  C_i(y,\phi)
  =\phaseavg{-u_z'u_y'I_i}
  =\phaseavg{u_z'u_r'I_i},
  \qquad i=2,4,
  \label{eq:quadrant-contribution}
\end{equation}
where $y=R-r$ is measured from the wall towards the pipe centre, so $u_y'=-u_r'$. The indicator $I_2$ is unity for Q2 ejections ($u_z'<0$, $u_y'>0$), and $I_4$ is unity for Q4 sweeps ($u_z'>0$, $u_y'<0$); both vanish otherwise. The inner-scaled contributions are $C_i^+=C_i/\utauzero^2$. The relation $C_i=P(Q_i)\phaseavg{-u_z'u_y'\mid Q_i}$ shows that each contribution combines event probability with conditional event intensity. The wall-normal location where $C_4/C_2=1$ defines the crossover position $\ycross(\phi)$, abbreviated to $\ycross$.

In statistically steady channel flow, Q4 sweeps dominate near the wall and Q2 ejections farther out, with comparable contributions at $y^+\approx12$ \citep{kim1987_turbulence_statistics}. Before boundary-layer transition, \citet{nolan2010_quadrant_transition} observed a wallward shift of the off-wall turbulent-kinetic-energy-production peak, with outer Q2 events associated with low-speed streaks and near-wall Q4 events strengthening. These observations provide reference points for the radial quadrant response considered here.

Motivated by the statistical characterisation of transition in pulsating flow \citep{taylor2024_turbulent_transient}, we distinguish the local instability point $\phi_{i,\mathrm{onset}}(y)$, defined in a phase-averaged statistical sense, from the subsequent discrete peak phase of each quadrant contribution. For each contribution, $C_{i,\min}^+$ and $C_{i,\max}^+$ denote the minimum and maximum values between deceleration onset and its discrete peak phase. The threshold is defined as
\begin{equation}
  C_{i,\mathrm{thr}}^+(y)
  =C_{i,\min}^+(y)
  +0.075\left[C_{i,\max}^+(y)-C_{i,\min}^+(y)\right],
  \qquad i=2,4.
  \label{eq:local-instability-threshold}
\end{equation}
Within this interval, $\phi_{i,\mathrm{onset}}(y)$ is the first sampled phase at which $C_i^+$ reaches or exceeds the threshold and increases over both the preceding and following phase intervals.
Comparison with the spectral and cospectral timing discussed in \S\ref{sec:spectra} guides the choice of 0.075 and supports interpreting the quadrant-response onset as a local instability point. Varying the threshold fraction between 0.05 and 0.10 changes the onset by at most one sampled phase relative to the 0.075 criterion. Onset phases at neighbouring wall distances may nevertheless exchange order or coincide.

During stage I, the radial distribution of $C_4/C_2$ changes little in either case. Both contributions remain approximately frozen near the wall in BP3496 but recover slowly near the end of acceleration in NP3496.

Figure~\ref{fig:q4-q2-ratio} shows that, during stage II, $C_4/C_2$ rises sharply in the viscous sublayer, more strongly in BP3496 than in NP3496, and peaks near the drag minimum. Simultaneously, the ratio decreases outside $\ycross$, indicating a larger relative Q2 contribution. Near the wall, the recovery of $C_2$ lags that of $C_4$ by a longer interval in BP3496 than in NP3496 (figure~\ref{fig:q2-q4-contributions}). Since the absolute stress contributions very near the wall are small, momentum-transport intensity must be assessed jointly from $C_2$, $C_4$ and the shear cospectrum. At $y^+<10$, $C_4$ increases more than $C_2$; near the initial spectral peak, the two contributions begin to rise at similar times, but $C_4$ grows faster.

The principal response regions show concurrent increases in relative intermediate- and small-scale energy and shear contribution at PHS-14--PHS-15 in BP3496 (figure~\ref{fig:normalised-spectra}(c)) and PHS-12--PHS-13 in NP3496, corroborating the local instability points identified there. Varicose deformation and breakdown of near-wall high-speed streaks accompany the Q4 increase, whereas sinuous deformation and breakdown of outer-buffer-layer low-speed streaks accompany the increasing relative Q2 contribution. As the near-wall Q4 influence becomes established farther out, $\ycross$ moves outwards.

By the drag minimum, $\ycross$ in BP3496 is already close to its outermost position. It moves only slightly farther out during early stage~III, reaching $y^+\approx17$ at PHS-21 before turning wallwards. In NP3496, its outermost location is $y^+\approx12$ at PHS-14, coinciding with the drag minimum. Near the wall, $C_4/C_2$ decreases towards the drag-recovery maximum as $C_2$ continues to grow while the growth of $C_4$ slows and then reverses. Farther out, delayed growth of $C_4$ raises the ratio near the drag-recovery maximum.

Table~\ref{tab:quadrant-timing} summarises the local instability points, peak phases and onset-to-peak intervals. Both contributions reach their peaks earlier and have shorter onset-to-peak intervals in NP3496 than in BP3496, although the radial peak-phase spans are comparable. By their respective drag-recovery maxima, both contributions have reached their sampled peaks over $y^+\approx2.2$--30.6 in BP3496 and $y^+\approx2.2$--20.8 in NP3496; the Q4 contribution farther out peaks later. The narrower range in NP3496 is consistent with the more restricted wall-normal extent of streak elongation identified in \S\ref{sec:spectra}.

\begin{table}
  \centering
  \small
  \setlength{\tabcolsep}{2pt}
  \begin{tabular}{@{}l*{6}{c}@{\hspace{8pt}}*{6}{c}@{}}
    \toprule
    & \multicolumn{6}{c}{BP3496} & \multicolumn{6}{c}{NP3496} \\
    \cmidrule(lr){2-7}\cmidrule(lr){8-13}
    $y^+$ & 2.2 & 7.8 & 14.2 & 20.8 & 30.6 & 60.7 & 2.2 & 7.1 & 11.1 & 20.8 & 30.6 & 60.7 \\
    \midrule
    Q2 onset & 17 & 17 & 15 & 14 & 14 & 16 & 14 & 13 & 12 & 12 & 12 & 13 \\
    Q2 peak & 26 & 26 & 26 & 25 & 24 & 26 & 19 & 19 & 19 & 18 & 18 & 20 \\
    Q2 growth interval & 9 & 9 & 11 & 11 & 10 & 10 & 5 & 6 & 7 & 6 & 6 & 7 \\
    \addlinespace
    Q4 onset & 16 & 15 & 15 & 15 & 15 & 18 & 13 & 12 & 12 & 12 & 13 & 15 \\
    Q4 peak & 24 & 24 & 24 & 25 & 26 & 28 & 18 & 18 & 18 & 19 & 20 & 22 \\
    Q4 growth interval & 8 & 9 & 9 & 10 & 11 & 10 & 5 & 6 & 6 & 7 & 7 & 7 \\
    \bottomrule
  \end{tabular}
  \caption{Local instability points, discrete peak phases and onset-to-peak intervals
  of the Q2 and Q4 shear contributions at the wall distances sampled
  in figure~\ref{fig:q2-q4-contributions}. Onset phases follow the
  criterion described in the text and equation~\eqref{eq:local-instability-threshold}.
  Onset and peak values are phase indices, and growth intervals
  are the differences between them.}
  \label{tab:quadrant-timing}
\end{table}

The earlier response in NP3496 is consistent with its higher turbulence level and vortex density at deceleration onset (figures~\ref{fig:vortical-structures}(a,e) and~\ref{fig:energy-spectra}(b,e)). Its narrower streak-elongation region and shorter radial adjustment distance may also contribute to the shorter growth intervals. Near-wall high-speed-streak instability and Q4 amplification accompany its earlier drag minimum; outer-buffer-layer low-speed-streak instability and delayed near-wall Q2 establishment accompany its earlier recovery maximum.

During stage IV, $C_4/C_2$ decreases slightly further near the wall but rises farther out as $C_2$ decays and $C_4$ peaks later. During the plateau in BP3496, both contributions decay approximately synchronously, while the ratio approaches its baseline distribution and $\ycross$ moves slowly outwards (figures~\ref{fig:q4-q2-ratio} and~\ref{fig:q2-q4-contributions}).

The local instability points $\phi_{i,\mathrm{onset}}(y)$ and peak phases distinguish how the two forms of streak breakdown and their shear contributions become established (figure~\ref{fig:q2-q4-contributions} and table~\ref{tab:quadrant-timing}). For $C_4$, the local instability points are nearly simultaneous across the viscous sublayer and buffer layer, whereas the peaks occur successively from the wall outwards. The shorter near-wall onset-to-peak interval may reflect the larger local mean-velocity gradient. This \downtop\ sequence, associated with predominantly varicose \innermode\ instability of near-wall high-speed streaks, describes the successive outward completion of breakdown and establishment of the Q4 contribution, rather than outward propagation of the instability onset. Some low-speed streaks also undergo varicose breakdown.

By contrast, in both BP3496 and NP3496, $C_2$ first increases in the outer buffer layer, with $\phi_{2,\mathrm{onset}}(y)$ progressively delayed towards the wall. Together with the streak evidence in \S\ref{sec:structures} and the spectra and cospectra in \S\ref{sec:spectra}, this ordering supports wallward transmission of low-speed-streak instability. The \topdown\ sequence associated with predominantly sinuous \outermode\ instability therefore primarily reflects the radial ordering of local instability onset. The interval from the local instability point to the peak is nevertheless shorter in the sampled viscous sublayer than at $y^+\approx20.8$--30.6, partly offsetting the delayed near-wall onset. The two processes overlap during the middle and late stages of deceleration and jointly restore multiscale shear transport.

\section{Mean-flow and turbulence dynamics}
\label{sec:turbulence-dynamics}

\subsection{Phase-averaged streamwise velocity and mean momentum transport}
\label{sec:momentum}

\begin{figure}
  \centering
  \projectfigure{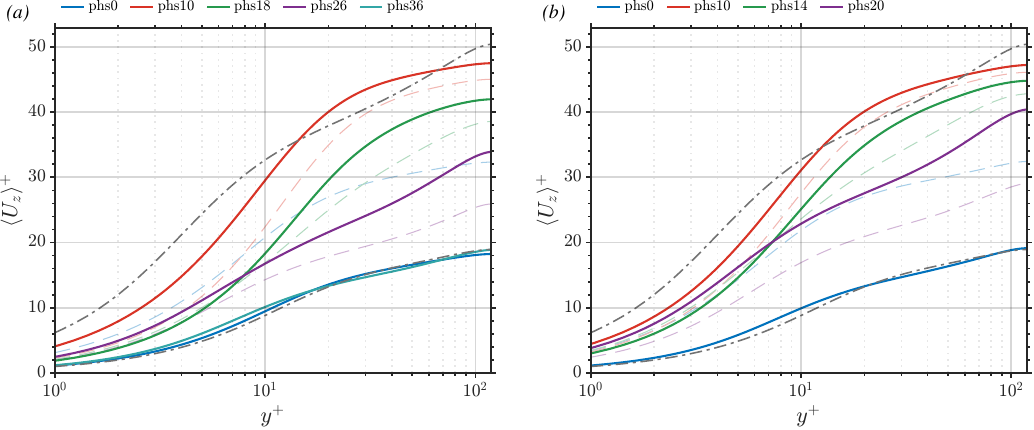}
  \caption{Phase-averaged streamwise-velocity profiles $\phaseavg{U_z}^+$ for (a) case BP3496 and (b) case NP3496. Coloured solid lines denote key phases and pale dashed lines denote intermediate phases. The key phases are PHS-0, 10, 18, 26 and 36 for BP3496, and PHS-0, 10, 14 and 20 for NP3496. Grey dash-dotted lines are the steady references at $\Reb=3400$ and 9600.}
  \label{fig:mean-velocity}
\end{figure}

\begin{figure}
  \centering
  \projectfigure{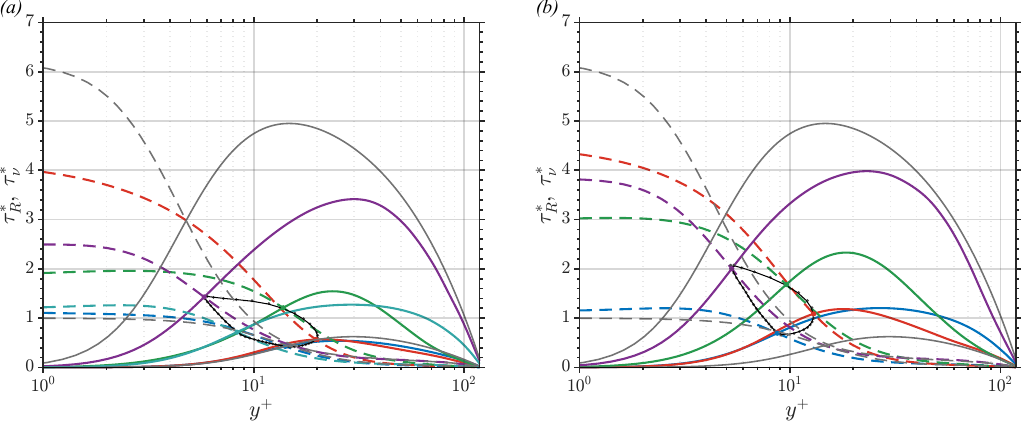}
  \caption{Local Reynolds and viscous shear stresses, $\tauR^*$ (solid) and $\tauv^*$ (dashed), and their crossover locations in (a) case BP3496 and (b) case NP3496. Key phases follow figure~\ref{fig:mean-velocity}. Grey circles, black solid lines and coloured filled circles denote the interpolated crossover at every phase, its smoothed trajectory and its value at the key phases, respectively. Grey reference curves correspond to $\Reb=3400$ and 9600.}
  \label{fig:local-shear-stresses}
\end{figure}

We now examine the phase-averaged velocity, Reynolds stresses and stress-budget terms to interpret the kinematic observations in terms of momentum and energy dynamics. Figures~\ref{fig:mean-velocity}--\ref{fig:transverse-normal-stresses} summarise the evolution of the velocity and stress profiles in BP3496 and NP3496 over a complete forcing cycle. The corresponding lower-Reynolds-number cases, BP3470 and NP3470, show similar stage-wise evolution. Key phases are highlighted in saturated colours, while intermediate profiles use paler shades of the colour assigned to the preceding key phase. Here, $\tauR=\rho\phaseavg{u_r'u_z'}$ and $\tauv=-\rho\nu\,\partial\phaseavg{U_z}/\partial r$ denote the Reynolds and viscous shear stresses, respectively.

Figure~\ref{fig:mean-velocity} shows that the streamwise-velocity profiles become increasingly plug-like during acceleration: most of the velocity increment occurs in the pipe core, with a smaller increase near the wall. From the onset of deceleration to the drag minimum, $\Ub$ has fallen only partly, whereas the wall drag drops rapidly. The core retains a high velocity and the near-wall gradient relaxes, reducing the drag well below the quasi-steady value at the same instantaneous flow rate. $\tauR$ has not yet fully recovered (figure~\ref{fig:local-shear-stresses}), demonstrating that wallward momentum transport lags the change in flow rate.

Between the drag minimum and drag-recovery maximum, $\Ub$ continues to fall, while the near-wall gradient and drag recover together. The concurrent increase in $\tauR$ links drag recovery to the re-establishment of turbulent wallward momentum transport; \S\ref{sec:quadrant} gives the timing of the quadrant contributions. From the drag-recovery maximum to the end of deceleration, the turbulence level remains above the steady reference at the same instantaneous Reynolds number. The $\tauR$--$\tauv$ crossover also remains closer to the wall. The total drag nevertheless approaches the quasi-steady curve. Section~\ref{sec:fik} explains this difference through the integral balance. Profile relaxation continues during the plateau in BP3496 and BP3470.

The profiles in figure~\ref{fig:local-shear-stresses} show that $\tauR$ remains approximately frozen or continues to decay during acceleration, whereas $\tauv$ grows rapidly in the viscous sublayer and spreads outwards. The near-wall viscous response therefore precedes the turbulent adjustment. Once deceleration begins, $\tauR$ recovers gradually; $\tauv$ falls sharply near the wall but continues to rise beyond the buffer layer, with a delay.
For the present cases, the ratio of the radial peak magnitude of $\tauR$ to that of $\tauv$ is approximately 0.7--0.8 near the drag minimum. Turbulent shear transport has thus recovered to the same order as viscous transport, but remains weaker than during the subsequent recovery stage. Both shear stresses decay after the drag-recovery maximum.

The crossover moves continuously outwards during acceleration, marking the expansion of the viscosity-dominated region. After deceleration begins, it reverses direction and moves towards the wall, reaches its innermost position around the drag-recovery maximum, and then moves outwards again. In BP3496, the reversal begins at approximately PHS-12; the crossover lies at $y^+\approx8.3$ at the end of deceleration and $y^+\approx14$ at the end of the plateau. In NP3496, both the wallward shift and the subsequent outward readjustment occur earlier, consistent with its earlier overall response.

\subsection{Evolution of the Reynolds stresses}
\label{sec:stress-evolution}

\begin{figure}
  \centering
  \projectfigure{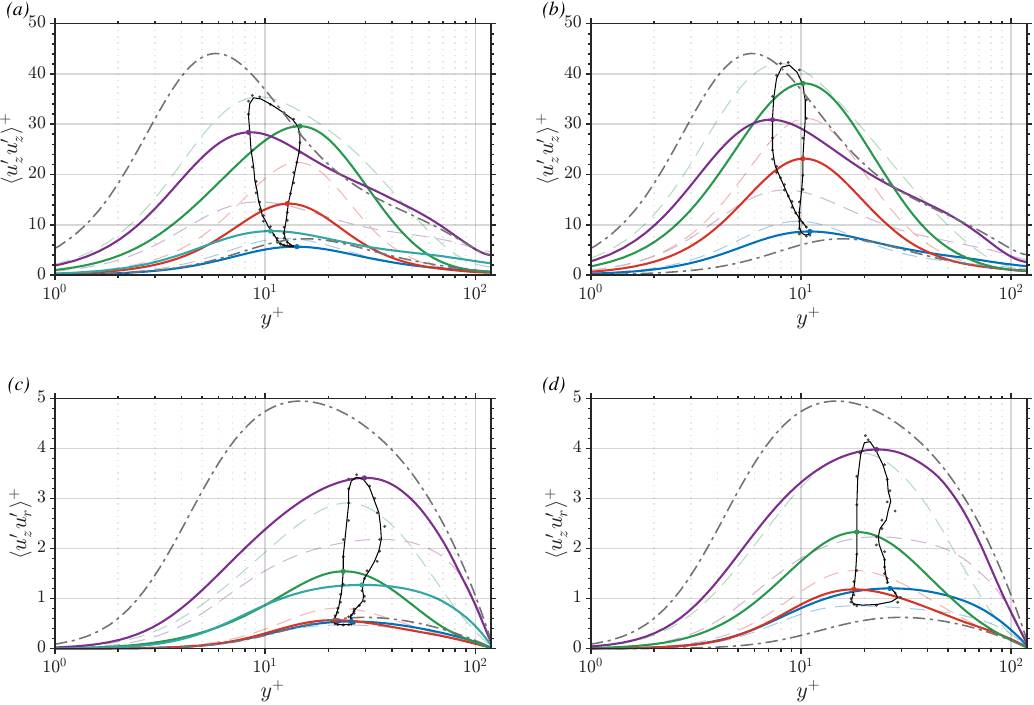}
  \caption{Phase-averaged Reynolds-stress profiles during the four stages in case BP3496 (a,c) and case NP3496 (b,d): (a,b) $\phaseavg{u_z'u_z'}^+$ and (c,d) $\phaseavg{u_r'u_z'}^+$. Coloured solid and pale dashed lines denote key and intermediate phases, respectively; the key phases follow figure~\ref{fig:mean-velocity}. Grey circles, black solid lines and coloured filled circles show the raw peak locations, smoothed peak trajectories and peak locations at the key phases, respectively. Grey dash-dotted lines are the steady references.}
  \label{fig:streamwise-shear-stresses}
\end{figure}

Figure~\ref{fig:streamwise-shear-stresses} compares the phase-averaged streamwise normal stress \(R_{zz}\) and Reynolds shear stress \(R_{rz}\), which characterise streamwise fluctuation energy and radial momentum transport, respectively. Here, \(R_{ij}=\phaseavg{u_i'u_j'}\), using the averaging operation defined in \S\ref{sec:averaging}. The plotted inner-scaled components are \(R_{ij}^+=R_{ij}/\utauzero^2\).

\begin{figure}
  \centering
  \projectfigure{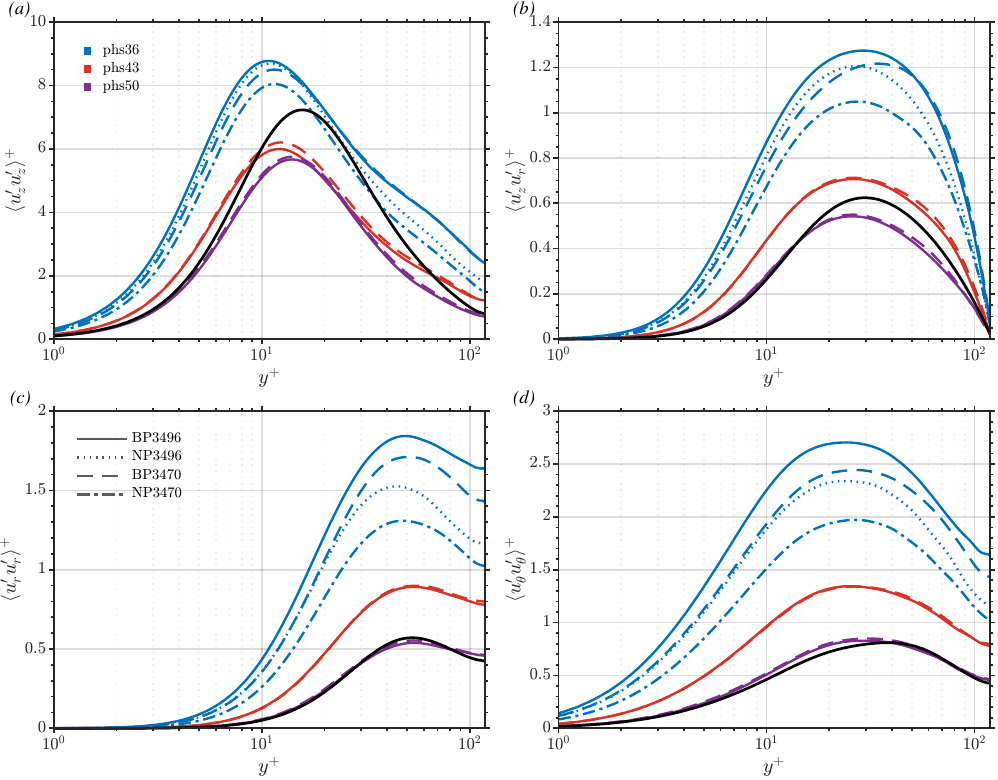}
  \caption{Phase-averaged Reynolds stresses during the plateau at the minimum $\Reb$ in cases BP3496 and BP3470, with PHS-0 of the following cycle in the no-plateau cases NP3496 and NP3470 (shown as PHS-36) used as the cycle-boundary reference: (a) $\phaseavg{u_z'u_z'}^+$, (b) $\phaseavg{u_r'u_z'}^+$, (c) $\phaseavg{u_r'u_r'}^+$ and (d) $\phaseavg{u_\theta'u_\theta'}^+$. Colours distinguish phases and line styles distinguish cases; the black solid line is the steady reference at $\Reb=3400$.}
  \label{fig:plateau-stresses}
\end{figure}

Early in acceleration, turbulence retains features of the preceding stage, consistent with the delayed adjustment reported for accelerating and decelerating pipe flows \citep{guerrero2021_accelerating_pipe,guerrero2023_decelerating_pipe}. In BP3496, the peak magnitude of \(R_{zz}\) changes little initially, retaining the low level reached at the end of the plateau (figures~\ref{fig:streamwise-shear-stresses}(a) and~\ref{fig:plateau-stresses}(a)). In NP3496, the decay of \(R_{zz}\) inherited from the preceding deceleration persists across the radial range into acceleration, corresponding to the residual-decay stage identified by \citet{taylor2024_turbulent_transient}. During this initial interval, \(R_{rz}\) decreases over the full radial range in both cases, with an initial peak magnitude in NP3496 approximately twice that in BP3496.
Recovery subsequently begins near the wall in both cases. The peaks of \(R_{zz}\) and \(R_{rz}\) first shift wallwards and then outwards, with only a slight outward shift of the \(R_{rz}\) peak. The peak magnitude of \(R_{zz}\) recovers slowly in BP3496. Both components recover earlier in NP3496, where the response of \(R_{rz}\) is also stronger. These early wallward peak shifts are governed primarily by the respective production terms, as discussed in \S\ref{sec:stress-budgets} and \S\ref{sec:transition-dynamics}.

From the onset of deceleration to the drag minimum, \(R_{zz}\) increases over the full radial range and its peak moves slowly outwards within the buffer layer. Around the drag minimum, \(R_{zz}\) at its radial-profile peak exceeds the steady $\Reb=9600$ value at the same wall distance. At the minimum, the peak lies at $y^+\approx14$--15 in BP3496 and $y^+\approx10$--11 in NP3496. \(R_{rz}\) also increases over most of the radial range, but its recovery continues to lag that of \(R_{zz}\).

Both stress components continue to increase after the drag minimum. The peak magnitude of \(R_{zz}\) reaches its maximum approximately three to four phases before the drag-recovery maximum in both cases. Its peak location continues to move wallwards to $y^+\approx8$ in BP3496 and 7 in NP3496. These locations are close to those of the energy-spectrum peaks (figure~\ref{fig:energy-spectra}(h,k)). \(R_{rz}\) recovers over most of the radial range, while its peak gradually shifts away from the wall within the buffer layer, indicating that both paths contribute to turbulence recovery and the restoration of momentum transport (\S\ref{sec:quadrant}). \(R_{rz}\) begins to decline locally only during the final one or two phases before the recovery maximum.

Both components decay after the drag-recovery maximum. In the pipe core, \(R_{zz}\) continues to increase briefly owing to the delayed local response, even as it decreases over most of the radial range. The peak of \(R_{zz}\) then shifts away from the wall more rapidly. The decrease in \(R_{rz}\) spreads to both sides of its peak; the peak first moves rapidly outwards and then gradually returns towards the wall. The initial outward shift accompanies the widespread presence of intermediate- and small-scale shear contributions in the outer-region cospectrum (figure~\ref{fig:shear-cospectra}(h,k,i,l)). \(R_{rz}\) decreases over the full radial range until the end of deceleration. This behaviour is consistent with decay after the outer-region shear contribution has become established.

\begin{figure}
  \centering
  \projectfigure{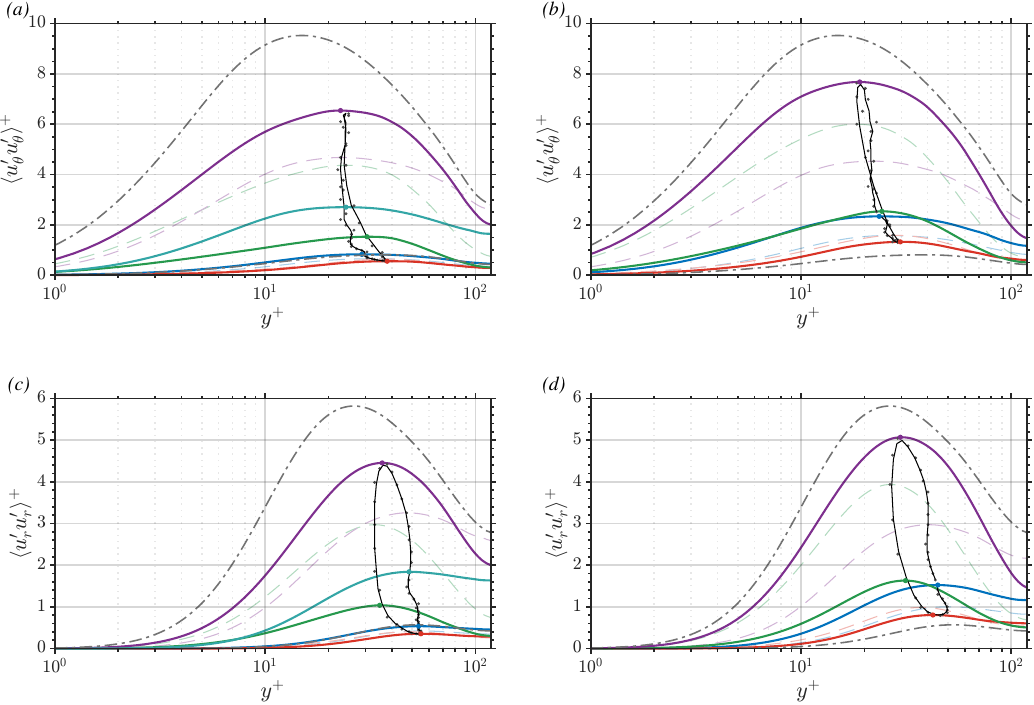}
  \caption{Phase-averaged transverse Reynolds-stress profiles during the four stages in case BP3496 (a,c) and case NP3496 (b,d): (a,b) $\phaseavg{u_\theta'u_\theta'}^+$ and (c,d) $\phaseavg{u_r'u_r'}^+$. Line styles, peak markers and steady references follow figure~\ref{fig:streamwise-shear-stresses}.}
  \label{fig:transverse-normal-stresses}
\end{figure}

Figure~\ref{fig:transverse-normal-stresses} compares the azimuthal normal stress \(R_{\theta\theta}\) and radial normal stress \(R_{rr}\), showing their common stage-wise redistribution of transverse energy and their distinct radial responses.
During acceleration, both transverse normal stresses decay overall and their peaks move outwards, indicating suppression of near-wall transverse fluctuations. This decay is consistent with the decrease in the number and density of vortical structures (figure~\ref{fig:vortical-structures}(a,e)). In NP3496, \(R_{rr}\) is initially markedly higher than in BP3496, with local recovery and the wallward shift of its peak occurring earlier.

From deceleration onset to the drag-recovery maximum, the radial-profile peak magnitudes of both transverse normal stresses increase. Recovery is initially slow and accelerates before the drag minimum, with the largest increase occurring during the rapid-transition stage between the two extrema. Before the drag minimum, the \(R_{rr}\) peak shifts wallwards from $y^+\approx55$ to 35 in BP3496 and from 42 to 31 in NP3496. Between the two extrema, it reaches its smallest wall distance, at approximately 33 and 26, respectively. The proximity of these locations to the outer buffer layer suggests a spatial correspondence with the region where outer-mode instability of low-speed streaks first develops (\S\ref{sec:structures}). The \(R_{\theta\theta}\) peak likewise moves wallwards before the drag minimum, from approximately 38 to 31 in BP3496 and from 30 to 24 in NP3496, subsequently reaching approximately 23 and 18, respectively, during rapid transition.

The peak magnitudes level off around the recovery maximum. The \(R_{rr}\) peak begins to move away from the wall during the final phases before the recovery maximum and continues in this direction afterwards. This readjustment may reflect the combined influence of the two transition paths at larger wall distances, consistent with the \(R_{rz}\) peak evolution and the quadrant analysis (\S\ref{sec:quadrant}). The \(R_{\theta\theta}\) peak remains closer to the wall than the \(R_{rr}\) peak, and its position varies non-monotonically around and after the recovery maximum. The spatial association of high- and low-speed-streak breakdown (\S\ref{sec:structures}) is consistent with interactions between the Q2- and Q4-related responses within their overlap region. These interactions may redistribute turbulent energy into azimuthal fluctuations, contributing to the rapid growth and more wallward concentration of \(R_{\theta\theta}\) during transition and to subsequent variations in its peak location. Both stresses decay over the full radial range after the recovery maximum, while the \(R_{\theta\theta}\) peak subsequently remains within a comparatively narrow radial band.

Figure~\ref{fig:plateau-stresses} summarises the radial evolution of all four Reynolds-stress components during the minimum-Reynolds-number plateaux in BP3496 and BP3470. The common end of deceleration in the no-plateau cases NP3496 and NP3470 provides a cycle-boundary reference for comparing the plateau response across Reynolds-number ranges.

At the common end of deceleration, the \(R_{zz}\) profiles in BP3496 and NP3496 are already similar in the buffer layer, with both peaks at $y^+\approx11$. \(R_{rz}\) responds later, so its profiles in BP3496 and NP3496 are not yet as similar. The lag, and hence the inter-case difference, is even more pronounced for the transverse normal stresses \citep{liu2024_time_delayed}. NP3496 generally exhibits lower stress levels than BP3496, consistent with its earlier drag-recovery maximum and the longer interval of subsequent decay. Cases BP3470 and NP3470 show the same tendency, but the inter-case differences are larger at the end of deceleration. This is consistent with their later drag-recovery maxima, which leave less time for subsequent decay.

All four Reynolds-stress components continue to decay during the plateaux in BP3496 and BP3470. The peak of \(R_{zz}\) moves further outwards and its magnitude undershoots the steady $\Reb=3400$ reference. The peaks of \(R_{\theta\theta}\) and \(R_{rr}\) also shift outwards overall, whereas the \(R_{rz}\) peak moves towards the wall. The stress profiles in BP3496 and BP3470 are nearly identical at the end of the plateau. Within the present parameter range, the plateau therefore weakens differences inherited from the preceding acceleration--deceleration segment.

Turbulence is already decaying during late deceleration, and this decay continues into the plateau. This differs from the approximately frozen turbulence over most of the pipe cross-section during the initial stage of rapid deceleration from a steady turbulent state \citep{guerrero2023_decelerating_pipe}.

The plateau therefore lowers the turbulence level inherited by the next acceleration stage. Further suppression during acceleration brings the transverse stresses below the steady $\Reb=3400$ reference. Without a plateau, NP3496 and NP3470 retain larger shear stress and transverse fluctuations and consequently develop energy accumulation, instability and drag recovery earlier. Too short a plateau may not reset the initial state fully, whereas an excessively long plateau may reduce transport efficiency per cycle or allow turbulence to recover. This mechanistic hypothesis offers a possible explanation for the optimal plateau duration reported by \citet{scarselli2023_cardiac_cycle}.

\subsection{Streamwise-normal-stress budget}
\label{sec:stress-budgets}

\begin{figure}
  \centering
  \projectfigure{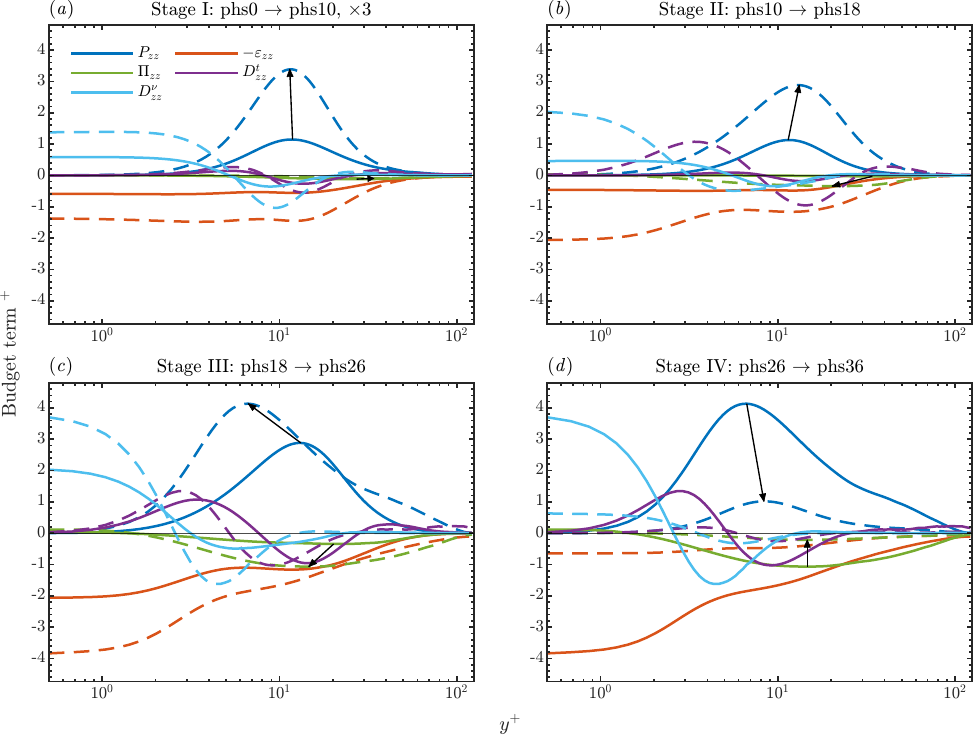}
  \caption{Streamwise-normal-stress budget in BP3496 during (a) acceleration, (b) deceleration towards the drag minimum, (c) drag recovery and (d) turbulence decay. Solid and dashed lines denote the earlier and later profiles, respectively. Arrows connect corresponding extrema of \(P_{zz}\) and \(\Pi_{zz}\). All terms in (a) are multiplied by three.}
  \label{fig:uzz-budget}
\end{figure}

To interpret the observed evolution of streaks and the streamwise normal stress, we examine the phase-averaged transport equation for \(R_{zz}\):
\begin{equation}
  \begin{aligned}
  \frac{\partial\phaseavg{u_z'u_z'}}{\partial t}
  ={}&\underbrace{-2\phaseavg{u_z'u_r'}
       \frac{\partial\phaseavg{U_z}}{\partial r}}_{P_{zz}}
     \underbrace{-2\nu\phaseavg{\boldsymbol{\nabla}u_z'
       \boldsymbol{\cdot}\boldsymbol{\nabla}u_z'}}_{-\varepsilon_{zz}}\\
   &+\underbrace{\frac{\nu}{r}\frac{\partial}{\partial r}
       \left(r\frac{\partial\phaseavg{u_z'u_z'}}{\partial r}\right)}
       _{D_{zz}^{\nu}}
     \underbrace{-\frac{1}{r}\frac{\partial}{\partial r}
       \left(r\phaseavg{u_r'u_z'u_z'}\right)}_{D_{zz}^{t}}
     +\underbrace{\frac{2}{\rho}\phaseavg{p'
       \frac{\partial u_z'}{\partial z}}}_{\Pi_{zz}}.
  \end{aligned}
  \label{eq:uzz-budget}
\end{equation}
Figure~\ref{fig:uzz-budget} shows the individual budget terms at representative phases near the stage boundaries.
For the present simulations, budget terms carrying a superscript \(+\) are normalised by \(\utauzero^4/\nu\).
With the unsteady term included, the BP3496 streamwise-normal-stress budget has a mean relative residual of $0.63\%$ over $0.5\leq y^+\leq100$, where the absolute imbalance is normalised locally by the sum of the absolute unsteady and budget terms before averaging.

Production \(P_{zz}\) depends jointly on the radial mean-velocity gradient \(\partial\phaseavg{U_z}/\partial r\) and the Reynolds shear stress \(R_{rz}\). The latter vanishes at the wall and pipe centre, while the magnitude of the mean-velocity gradient decreases rapidly away from the wall; the production peak therefore lies between the wall and the peak of \(R_{rz}\). The correspondence between the evolution of \(P_{zz}\) and \(R_{zz}\) supports a production-based interpretation of streamwise-energy growth and its radial variation.

During early acceleration, the magnitude of \(P_{zz}\) changes little, while its peak shifts wallwards as the near-wall mean-velocity gradient strengthens. The subsequent weak increase in \(R_{rz}\) (figure~\ref{fig:streamwise-shear-stresses}(c,d)) strengthens its radial overlap with the mean gradient. Production then grows appreciably and its peak moves gradually outwards, with peak locations at \(y^+\approx10\)--12 in BP3496 and \(y^+\approx8\)--9 in NP3496.

The weak increase in shear transport is consistent with the small increase in the shear-cospectrum peak as it shifts towards longer wavelengths (figure~\ref{fig:shear-cospectra}(a,b,d,e)). As \(R_{zz}\) recovers while the transverse normal stresses continue to decay, newly generated fluctuation energy accumulates primarily in the streamwise component, while transverse redistribution and radial momentum transport lag behind. Together, the production and stress responses explain the initial wallward shift and subsequent outward adjustment of the peaks in both the premultiplied streamwise spectrum and the \(R_{zz}\) profile.

Once deceleration begins, the near-wall mean-velocity gradient decreases rapidly but remains comparatively large within the buffer layer. \(R_{rz}\) and \(P_{zz}\) continue to grow. The outward shift of the production peak reflects the \downtop\ establishment of Q4-related shear transport (\S\ref{sec:quadrant}). Before the drag minimum, the magnitude of \(\Pi_{zz}\) increases slowly, and transverse redistribution and radial momentum transport continue to lag.

After the drag minimum, production and the magnitude of \(\Pi_{zz}\) increase markedly, accompanied by more rapid growth of the normal stresses. With the delayed near-wall amplification of Q2 during the \topdown\ process, the production peak reverses towards the wall and its magnitude begins to fall approximately two phases before the drag-recovery maximum. Recovery of the near-wall mean gradient and the increase in \(R_{rz}\) maintain this near-wall concentration, while the gradient has already weakened in the buffer and outer regions. Very-long-wavelength energy persists alongside markedly increasing intermediate- and small-scale energy. The spectral peak initially moves wallwards, with a transient increase in the dominant streamwise wavelength in BP3496, before shifting to shorter wavelengths near the recovery maximum. The magnitude of \(\Pi_{zz}\) reaches a maximum near the drag-recovery maximum and subsequently decreases, consistent with the delayed transverse redistribution of streamwise fluctuation energy.

After the drag-recovery maximum, \(R_{rz}\) and the mean-velocity gradient weaken together. Production, dissipation and diffusion weaken overall, while \(\Pi_{zz}\) decreases in magnitude and the production peak shifts outwards. The spectral peak likewise weakens and shifts outwards towards longer wavelengths (figure~\ref{fig:energy-spectra}(h,k,i,l)).
During the plateau, this decay and outward adjustment continue, with the peak of \(|\Pi_{zz}|\) moving from \(y^+\approx15\) to 30.

\subsection{Transition dynamics}
\label{sec:transition-dynamics}

\begin{figure}
  \centering
  \projectfigure{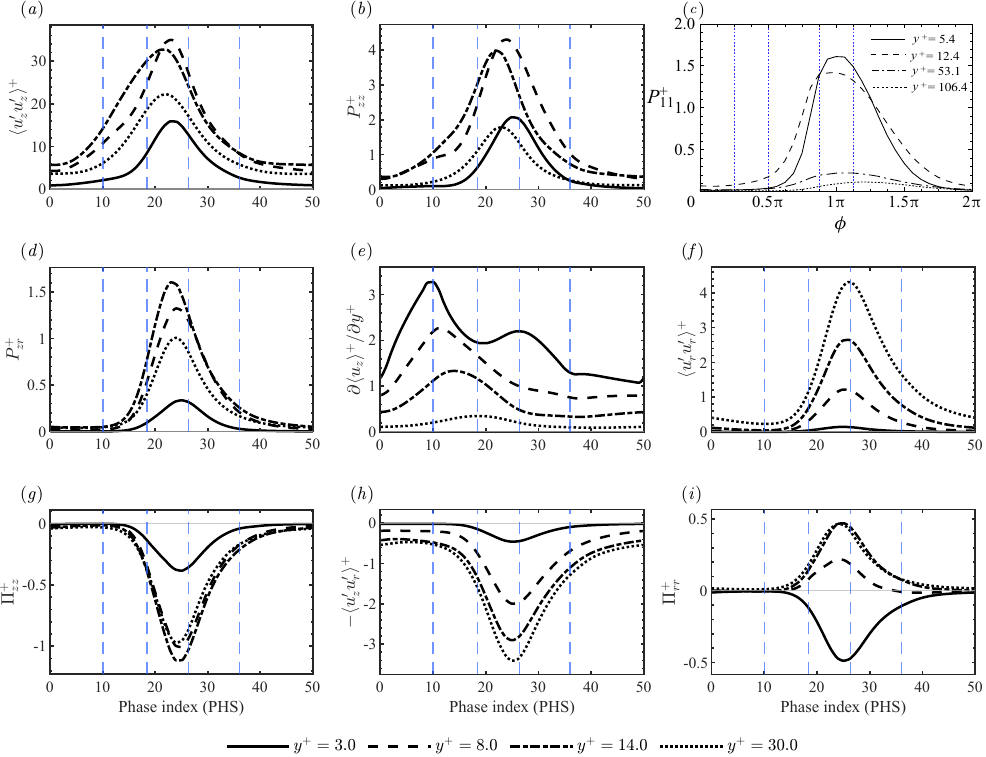}
  \caption{Phase evolution at four wall distances in case BP3496: (a) $R_{zz}^+$, (b) $P_{zz}^+$, (c) $P_{11}^+$, (d) $P_{zr}^+$, (e) $\partial\phaseavg{U_z}^+/\partial y^+$, (f) $R_{rr}^+$, (g) $\Pi_{zz}^+$, (h) $-R_{rz}^+$ and (i) $\Pi_{rr}^+$. Wall distances are indicated in the legends. Except in (c), blue vertical dashed lines mark the end of acceleration, the drag minimum, the drag-recovery maximum and the end of deceleration. Panel (c), showing case L26A05 with the phase coordinate \(\phi\), is adapted from figure~11(e) of \citet{taylor2024_turbulent_transient} under the \href{https://creativecommons.org/licenses/by/4.0/}{Creative Commons Attribution 4.0 licence}.}
  \label{fig:transition-case-a}
\end{figure}

\begin{figure}
  \centering
  \projectfigure{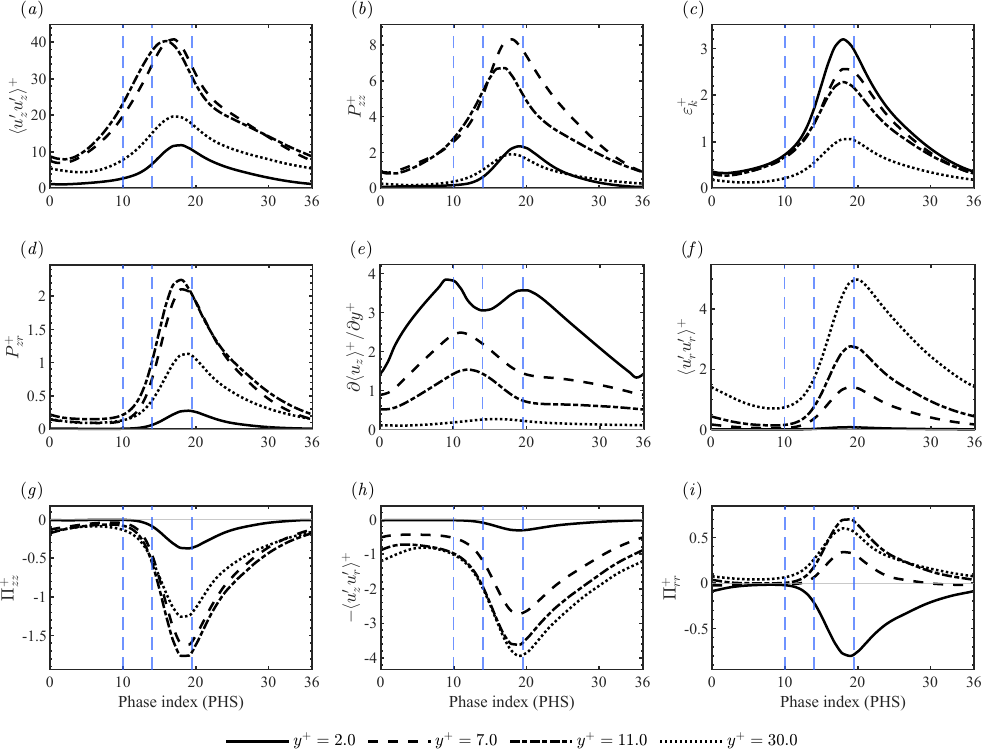}
  \caption{Phase evolution at four wall distances in case NP3496: (a) $R_{zz}^+$, (b) $P_{zz}^+$, (c) $\varepsilon_k^+$, (d) $P_{zr}^+$, (e) $\partial\phaseavg{U_z}^+/\partial y^+$, (f) $R_{rr}^+$, (g) $\Pi_{zz}^+$, (h) $-R_{rz}^+$ and (i) $\Pi_{rr}^+$. Here, $\varepsilon_k$ denotes the dissipation rate of turbulent kinetic energy. Wall distances are indicated in the legend. Blue vertical dashed lines mark the end of acceleration, the drag minimum and the drag-recovery maximum.}
  \label{fig:transition-case-c}
\end{figure}

Figures~\ref{fig:transition-case-a} and \ref{fig:transition-case-c} compare the stress and transport histories to interpret the onset and completion of transition. The selected wall distances, \(y^+=3\), 8, 14 and 30 in BP3496 and \(y^+=2\), 7, 11 and 30 in NP3496, are close to corresponding locations in figure~\ref{fig:q2-q4-contributions} and span the near-wall region, buffer layer and its outer edge.
The additional terms used below are the shear-stress production \(P_{zr}=-R_{rr}\,\partial\phaseavg{U_z}/\partial r\) and the radial pressure--strain term \(\Pi_{rr}=(2/\rho)\phaseavg{p'\,\partial u_r'/\partial r}\).

The phase histories show that \(R_{zz}\) and \(P_{zz}\) recover before \(R_{rr}\), \(R_{rz}\), \(|\Pi_{zz}|\) and \(|\Pi_{rr}|\). During deceleration, \(P_{zr}\) and \(R_{rz}\) strengthen as \(\Pi_{zz}\) becomes more negative and the positive radial component \(\Pi_{rr}\) increases within the buffer layer. These responses support the delayed sequence ``streamwise energy input--pressure--strain redistribution--radial momentum transport''.
\(R_{zz}\) and \(P_{zz}\) peak at phases that vary with radial position, generally later near the wall. The other stresses and transport terms show weaker radial differences, while their decay begins more nearly simultaneously, further indicating that the redistribution stage is influenced by both transition paths.

We interpret the stress and transport responses within the four-stage drag framework introduced in \S\ref{sec:drag-energy}, using the turbulent--turbulent transient classification of \citet{taylor2024_turbulent_transient} for comparison. Using the full cycle period $T$ and $u_{\tau,\mathrm{cyc}}=\sqrt{\overline{\tau_w}/\rho}$, where the overbar denotes a cycle average, the Stokes length $\ell_{s,\mathrm{cyc}}^+=u_{\tau,\mathrm{cyc}}\sqrt{T/(\pi\nu)}$ is approximately 18 in both BP3496 and NP3496. This value lies numerically between their reported values 16 and 26, although their friction scale is based on steady flow at the cycle-averaged Reynolds number. Since the present flow remains turbulent throughout the cycle and is driven by a piecewise-linear waveform, the comparison is limited to the stage characteristics of their L26A05 case (\(A=0.5\)), shown in figure~\ref{fig:transition-case-a}(c). 

In L26A05, Taylor \& Seddighi identified pre-transition by the slow growth of streamwise-normal-stress production and approximately frozen wall-normal fluctuations near the wall. Transition in this case occurred during acceleration and was marked by an abrupt increase in the production growth rate, associated with low-speed-streak instability, breakdown and turbulent-spot growth. They also found that increasing the pulsation amplitude delays transition. In the present flow, the stronger acceleration postpones the response until deceleration and permits greater prior accumulation of streamwise energy. For the turbulence-decay stage, their criterion was the onset of decay in wall-normal and spanwise fluctuations near the wall, allowing for a delayed core response. Consistent with this criterion, \(R_{rr}\) at the sampled wall distances begins to decay around the present drag-recovery maximum (figures~\ref{fig:transition-case-a}(f) and~\ref{fig:transition-case-c}(f)). Stage~IV is therefore interpreted as a turbulence-decay stage, with the drag-recovery maximum providing its temporal boundary.

During early deceleration, the increase in \(R_{rz}\) is partly offset by the weakening mean-velocity gradient, and the change in the growth rate of \(P_{zz}\) varies with wall distance (figures~\ref{fig:transition-case-a}(b) and \ref{fig:transition-case-c}(b)). This variation is particularly relevant to low-speed-streak instability, which first develops in the outer buffer layer, where the production-growth change can be less distinct.

The radial variation in the production turning point reflects the relative changes in \(R_{rz}\). At the near-wall locations in figure~\ref{fig:transition-case-a}(b,e,h), the initial value of \(R_{rz}\) is very small and subsequently increases by more than an order of magnitude, in some cases by a factor of several tens. Farther out, the initial value of \(R_{rz}\) is higher and its relative increase remains below one order of magnitude despite a large absolute increment. Since the mean-velocity gradient at these locations remains below four times its initial value throughout the cycle, relative changes in \(R_{rz}\) modulate production more strongly near the wall. The growth-rate change therefore becomes progressively less distinct outwards.

This radial dependence makes the production-growth criterion less robust for identifying local transition onset at a single wall distance. Local transition is therefore taken to begin at the instability point $\phi_{i,\mathrm{onset}}(y)$ defined in \S\ref{sec:quadrant}. This point is extracted operationally from the onset of the quadrant response and corroborated by the phase-averaged spectra and shear cospectra.

The local instability point $\phi_{i,\mathrm{onset}}(y)$ marks the radius-dependent onset of transition, while the four-stage framework retains common temporal boundaries based on the forcing and wall-drag response (\S\ref{sec:drag-energy}). Stage~II extends from deceleration onset to the drag minimum and is characterised by drag undershoot and gradual turbulence recovery. Its beginning marks the overall change from suppression of transverse fluctuations and vortical activity to their gradual recovery (figures~\ref{fig:vortical-structures}(a,b,e,f) and~\ref{fig:transverse-normal-stresses}). Stage~III encompasses rapid transition and drag recovery from the drag minimum to the drag-recovery maximum, following the earlier onset of local instability in Stage~II.

Differences in the amplitude growth of high- and low-speed streaks during elongation (figure~\ref{fig:streak-structures}(a)) can be interpreted through the stress-production mechanism. The Q2 and Q4 contributions enter \(P_{zz}\) through \(R_{rz}\): Q4 dominates near the wall, whereas Q2 contributes more outside $\ycross$ (figure~\ref{fig:q4-q2-ratio}); under the local mean shear, this partition favours amplification of near-wall high-speed streaks and outer-buffer-layer low-speed streaks, respectively. Across the viscous sublayer and buffer layer, the earliest local instability points associated with Q4 and Q2 occur at similar phases, indicating that the two streak types first become unstable almost simultaneously (figure~\ref{fig:q2-q4-contributions}). The close timing of these earliest onsets, together with the spatial association of high- and low-speed-streak breakdown (\S\ref{sec:structures}), is consistent with coupled development of the two responses.

The local instability points of $C_2$ subsequently occur progressively later towards the wall. After outer-buffer-layer low-speed streaks become unstable, the enhanced turbulent shear supports transmission of the instability to radially adjacent regions. Once the \topdown\ process reaches the near-wall region, the initially weak local turbulent shear produces delayed growth of the low-speed-streak amplitude (figure~\ref{fig:streak-structures}(c--e)) and the associated $P_{zz}$.

For $C_4$, the radial ordering is expressed mainly through its peak phases. During the outward establishment of the \downtop\ process, turbulent shear remains weak; the local evolution is governed mainly by the large near-wall mean shear and its subsequent growth within the buffer layer. Consequently, the earlier growth and breakdown of near-wall high-speed streaks do not substantially alter the breakdown process of high-speed streaks farther from the wall. The larger initial turbulent shear farther out also makes the delayed growth of high-speed-streak amplitude and the associated $P_{zz}$ weaker and more readily obscured when the two paths overlap.

As the two shear contributions overlap, near-wall $C_4/C_2$ reaches its maximum close to the drag minimum. By the drag minimum, the crossover $\ycross$ has reached its outermost position in NP3496 and is already close to it in BP3496. The energy-spectrum peak and the peak of \(R_{zz}\) begin to move wallwards (figures~\ref{fig:energy-spectra}(c,f,g,j,h,k) and~\ref{fig:streamwise-shear-stresses}(a,b)). These changes develop successively around the drag minimum and are accompanied by increasing vortical activity, marked growth of \(R_{rr}\) and \(R_{\theta\theta}\), and stronger intermediate- and small-scale contributions to the streamwise spectrum and shear cospectrum (figures~\ref{fig:vortical-structures}(c,g), \ref{fig:transverse-normal-stresses} and~\ref{fig:shear-cospectra}(c,f,g,j,h,k)).

In NP3496, \(P_{zz}\), \(|\Pi_{zz}|\), \(|\Pi_{rr}|\) and \(R_{rz}\) increase and reach their maxima earlier. Across the sampled viscous sublayer and buffer layer, all $C_2$ contributions have peaked and entered decay before all $C_4$ contributions have done so (figure~\ref{fig:q2-q4-contributions}(c,k)); this compares the times by which each contribution has peaked across the region, rather than the ordering at every individual location. The longer local growth intervals in BP3496 also make the two paths easier to distinguish.

Under the present strong acceleration, streamwise energy input recovers ahead of transverse redistribution and radial momentum transport, allowing energy to accumulate during acceleration and early deceleration. The plateau extends the preceding decay and lowers the transverse fluctuations and \(R_{rz}\) inherited by the next acceleration stage (figure~\ref{fig:plateau-stresses}). Consistent with this interpretation, BP3496 exhibits later local instability points and longer onset-to-peak intervals than NP3496 (\S\ref{sec:quadrant}). Both the onset and completion of transition are therefore delayed in the plateau case, sustaining weak momentum transport for longer.

\begin{figure}
  \centering
  \projectfigure{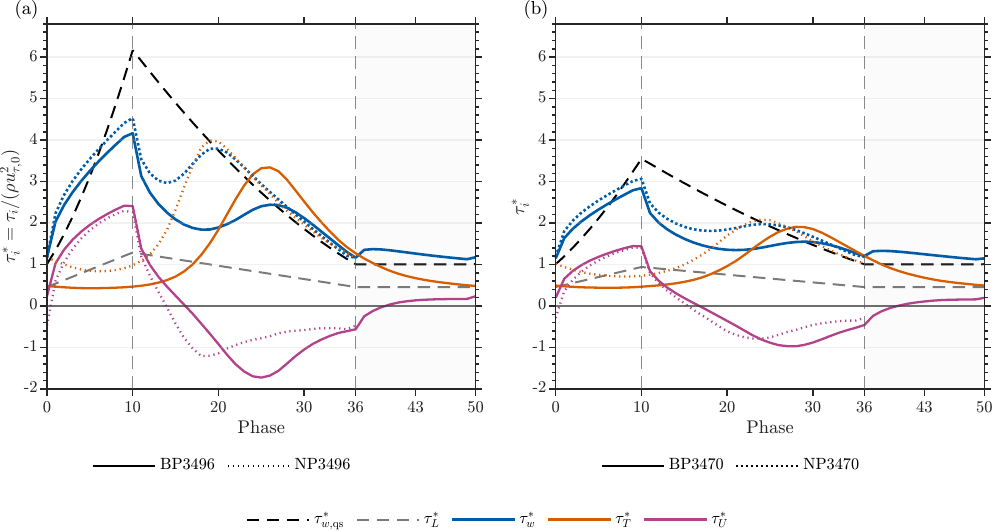}
  \caption{Normalised wall shear stress and unsteady FIK contributions
  for (a) BP3496 and NP3496 and (b) BP3470 and NP3470.
  Solid and dotted lines denote the plateau and no-plateau cases,
  respectively. Black and grey dashed curves denote the quasi-steady
  turbulent and laminar references shared within each case pair,
  respectively. Vertical dashed lines mark the ends of acceleration
  and deceleration; shading indicates the minimum-flow plateau.}
  \label{fig:fik-decomposition}
\end{figure}

By distinguishing local instability points from subsequent shear-contribution peaks, the present analysis resolves inward Q2-related development and outward establishment of the Q4 shear contribution. The production-based interpretation suggests that these pathways need not be specific to cardiac-cycle-inspired forcing. Their separation may depend on the initial turbulent state and the relative time scales of acceleration and turbulence adjustment. With weaker acceleration or a more energetic initial turbulent state, instability and shear recovery may develop earlier and overlap more strongly, making the distinct radial sequences less apparent in production histories alone.

Section~\ref{sec:fik} connects the delayed turbulence responses in the present cases to the wall-drag balance and examines their implications for drag reduction and pumping-energy savings.

\section{Unsteady FIK decomposition of wall friction}
\label{sec:fik}

The four-stage wall-drag response described in \S\ref{sec:drag-energy}
is interpreted using an unsteady FIK decomposition.
The pipe-flow form of the Fukagata--Iwamoto--Kasagi identity
\citep{fukagata2002_skin_friction,guerrero2021_accelerating_pipe}
is written directly in terms of wall shear stress, permitting comparison
across phases without a time-dependent dynamic-pressure normalisation:
\begin{equation}
  \begin{aligned}
    \tauw={}&
    \underbrace{\frac{4\rho\nu\Ub}{R}}_{\tauL}
    +\underbrace{
      \frac{4\rho}{R^3}\int_0^R
      r^2\phaseavg{u_r'u_z'}\,\mathrm{d}r
    }_{\tauT}\\[0.5ex]
    &+\underbrace{
      \tauw-\frac{4\rho}{R^3}\int_0^R r^2
      \left(
        \phaseavg{u_r'u_z'}
        -\nu\frac{\partial\phaseavg{U_z}}{\partial r}
      \right)\,\mathrm{d}r
    }_{\tauU}.
  \end{aligned}
  \label{eq:fik-wall-shear}
\end{equation}

The terms $\tauL$ and $\tauT$ denote the laminar contribution at the
instantaneous $\Ub$ and the radially weighted Reynolds-shear-stress
contribution, respectively. The unsteady contribution $\tauU$ measures
the integrated departure of total shear stress from the linear distribution of statistically steady,
fully developed flow. In the formulation of
\citet{guerrero2025_spectral_net_force}, the unsteady term is expressed
through radial gradients of normalised total shear stress; the present
dimensional form evaluates the corresponding residual from the
phase-averaged wall stress, mean velocity gradient and Reynolds shear
stress. This contribution is associated with the non-equilibrium effects
of pressure forcing and local inertia and vanishes in statistically
steady, fully developed flow.

Figure~\ref{fig:fik-decomposition} compares the normalised wall shear stress and FIK contributions for the plateau and no-plateau cases over the two Reynolds-number ranges. Each case is shown over one forcing cycle.

During stage I, $\tauT$ does not grow with the rapidly increasing $\Ub$, but remains nearly constant or decays by varying amounts across the cases. The rise in $\tauw$ is driven mainly by increasing $\tauL$ and positive $\tauU$, while the weak turbulent contribution keeps the acceleration-stage maximum below the quasi-steady prediction. The initial $\tauT$ is lower in BP3496 than in NP3496, and the plateau/no-plateau contrast is also evident in BP3470 and NP3470. At the end of acceleration, BP3496 and NP3496 have nearly identical $\tauL$ and similar $\tauU$, but $\tauT$ is higher in NP3496. The plateau thus lowers the turbulent contribution carried into the next acceleration stage and reduces the acceleration-stage drag maximum.

During stage II, $\tauT$ begins to recover, but the rapid decline of $\tauU$ into negative values dominates the undershoot of $\tauw$. The drag undershoot therefore reflects the non-equilibrium adjustment of the mean-velocity and total-shear-stress profiles after deceleration begins, rather than delayed turbulence recovery alone. Meanwhile, the streamwise normal stress grows rapidly, with the shear stress and radial momentum transport lagging behind (\S\ref{sec:transition-dynamics}). Following a brief adjustment, all four cases exhibit an approximately linear decline of $\tauU$ during early deceleration. Within each Reynolds-number range, the plateau case shows a slower early decline but reaches a more negative minimum; in BP3496, the decline also persists longer than in NP3496. The near-linear decline accompanies constant bulk deceleration but also reflects the non-equilibrium evolution of the velocity profile.

During stage III, the growth of $\tauT$ initially outweighs the decreases in $\tauL$ and $\tauU$, driving drag recovery. Once $\tauU$ begins to recover, it further supports the rise in wall stress as $\tauT$ approaches or passes its maximum. Compared with NP3496, the lower initial $\tauT$ in BP3496, the slower recovery of this contribution and the longer persistence of negative $\tauU$ explain its longer drag-decrease and recovery intervals (\S\ref{sec:drag-energy}). The larger $\tauT$ and less negative $\tauU$ in NP3496 jointly produce an earlier, higher recovery maximum. In panel (b), BP3470 and NP3470 exhibit the same sequence of undershoot and recovery through the evolving balance among $\tauL$, $\tauT$ and $\tauU$, with later drag extrema accompanying their lower Reynolds-number range and associated acceleration--deceleration rates.

During stage IV, $\tauT$ decays while $\tauU$ recovers, so compensation between them can produce similar total drag despite differences in the turbulent contribution.
At the common end of deceleration, BP3496 and NP3496 have similar mean-velocity and Reynolds-shear-stress profile shapes and nearly equal wall stress. The turbulent contribution remains approximately $8\,\%$ higher in BP3496, largely offset by a more negative unsteady contribution. Similar compensation has been reported during relaxation following rapid deceleration \citep{guerrero2023_decelerating_pipe}; here, it occurs while the bulk flow is still decelerating.

Compared with their no-plateau counterparts, BP3496 and BP3470 retain higher turbulent contributions during late deceleration, including after their drag-recovery maxima. These contributions decline more steeply near the end of deceleration and continue to decay through the plateau, consistent with the Reynolds-stress evolution in \S\ref{sec:stress-evolution}. Both plateau cases exhibit a brief drag recovery followed by a slow decline. During the plateau, $\tauL$ remains nearly constant, while the recovery of $\tauU$ partly offsets the decay of $\tauT$, producing a comparatively weak drag response. The weak variation in total drag therefore does not imply equilibrium. NP3496 enters the next acceleration stage immediately after deceleration and retains higher $\tauT$ and negative $\tauU$, whereas BP3496 ends its plateau with lower $\tauT$ and positive $\tauU$.

The unsteady FIK balance connects the transition dynamics in \S\ref{sec:transition-dynamics} to the drag response of the present cardiac-cycle-inspired waveforms. During acceleration, streamwise energy accumulates while \(\tauT\) responds only weakly to the increasing flow rate. During early deceleration, the rapid decline of \(\tauU\) into negative values dominates the wall-stress undershoot while \(\tauT\) begins to recover. The subsequent establishment of Q2- and Q4-related shear transport accompanies the recovery of \(\tauT\), while the later rise of \(\tauU\) further increases wall drag. Delayed onset and longer development intervals sustain weak turbulent transport, while the combined evolution of \(\tauL\), \(\tauT\) and \(\tauU\) determines the depth and duration of the low-drag window.

The plateau reinforces this delay by weakening the turbulence inherited by the next acceleration stage, lowering the initial \(\tauT\) and modifying the amplitude and duration of the \(\tauU\) response. Since cycle-averaged pumping power is proportional to \(\overline{\tauw\Ub}\), the energetic benefit depends on both the duration of low drag and its timing relative to \(\Ub\). In the present paired comparisons, the plateau cases yield higher \(DR\) and \(ES\) than their no-plateau counterparts, with larger improvements in the higher Reynolds-number pair (table~\ref{tab:case-discretisation}).

\section{Summary and conclusions}
\label{sec:conclusions}

We performed direct numerical simulations to examine the phase-dependent turbulence dynamics of four pulsatile pipe flows driven by cardiac-cycle-inspired waveforms, comparing plateau and no-plateau forcing over two Reynolds-number ranges. Detailed analysis of BP3496 and NP3496, which span the higher Reynolds-number range, relates structural evolution to momentum and energy dynamics throughout the forcing cycle. The response is organised into four stages defined by the forcing and wall-drag history. The flow remains turbulent throughout; transition denotes local streak instability and turbulent--turbulent reorganisation.

Stage I corresponds to strong acceleration, during which transverse fluctuations and vortical activity are suppressed. Following the inherited decay, streamwise production and normal stress recover ahead of pressure--strain redistribution and Reynolds shear stress, allowing energy to accumulate in elongated streaks at long streamwise scales. The integrated turbulent contribution $\tauT$ does not grow synchronously with the flow rate. Increasing $\tauL$ and positive $\tauU$ drive the rise in wall drag, whose acceleration-stage maximum remains below the quasi-steady prediction.

Stage II extends from the onset of deceleration to the drag minimum. Local instability develops: near-wall high-speed streaks exhibit predominantly varicose \innermode\ disturbances associated with Q4 sweeps, whereas outer-buffer-layer low-speed streaks exhibit predominantly sinuous \outermode\ disturbances associated with Q2 ejections. Long-scale energy persists alongside recovering intermediate- and small-scale contributions. Streamwise energy input continues while shear transport recovers and transverse redistribution still lags. Although $\tauT$ begins to increase, the rapid decline of $\tauU$ into negative values dominates the drag undershoot.

Stage III spans rapid transition and drag recovery between the drag minimum and recovery maximum, following the local instability initiated during stage II. As the two pathways overlap, vortical activity and pressure--strain redistribution intensify markedly, while multiscale momentum transport recovers rapidly. The streamwise-energy peak shifts wallwards as energy decays at its original location and near-wall low-speed streaks amplify, while the Reynolds-shear-stress peak moves outwards. Both transverse normal stresses grow rapidly. Drag recovery is driven primarily by growth of $\tauT$, which initially outweighs the decreases in $\tauL$ and $\tauU$.

Stage IV is characterised by turbulence decay between the drag-recovery maximum and the onset of the next acceleration stage. Vortical activity weakens first near the wall and later farther out, while the characteristic streak scale increases and the spectral peak moves outwards. Reynolds stresses and the principal terms of the streamwise-normal-stress budget decay overall but remain out of equilibrium. At the common end of deceleration, compensation between $\tauT$ and $\tauU$ can produce similar wall drag despite different turbulent shear contributions.

The transition exhibits distinct radial sequences of local instability onset and subsequent shear-contribution peaks. The $C_2$ local instability points occur progressively later towards the wall, but the subsequent onset-to-peak interval is shorter in the sampled viscous sublayer than in the outer buffer layer. The Q2 contribution peaks first in the outer buffer layer and later towards both the wall and the outer region. Together with the structural and spectral evidence, this timing supports the \topdown\ development of low-speed-streak instability.

The $C_4$ local instability points are nearly simultaneous across the viscous sublayer and buffer layer, whereas the peaks occur successively from the wall outwards. This \downtop\ sequence describes the order of high-speed-streak breakdown and shear-contribution peaks, rather than outward propagation of the instability onset. Varicose disturbances also occur in low-speed streaks; the modes are not exclusive to one streak type.

Differences in streak amplification during elongation can be interpreted through streamwise production, which depends jointly on mean shear and Reynolds shear stress. Under the local mean shear, Q4 dominance near the wall favours high-speed-streak amplification, whereas Q2 dominance in the outer buffer layer favours low-speed-streak amplification. The subsequent outward establishment of the Q4-related response is governed mainly by the large near-wall mean shear and its growth within the buffer layer. For the Q2-related response, enhanced turbulent shear following outer-buffer-layer low-speed-streak instability supports wallward development, while the initially weak near-wall turbulent shear delays low-speed-streak amplification.

The minimum-Reynolds-number plateau extends structural relaxation and the decay of all Reynolds-stress components. The next acceleration stage therefore inherits weaker transverse fluctuations and lower shear stress, with further suppression bringing the transverse stresses below the steady $\Reb=3400$ reference. In the detailed comparison, the plateau further delays local instability and lengthens the onset-to-peak intervals of the quadrant contributions, sustaining weak momentum transport for longer. Across the plateau cases, the FIK balance also shows a lower initial $\tauT$ and changes in the amplitude and duration of negative $\tauU$. The plateau therefore reinforces the delayed turbulence recovery associated with strong acceleration and prolongs the low-drag window.

The drag minimum and recovery maximum occur later over the lower Reynolds-number range, where the imposed acceleration and deceleration rates are also lower. All four cases exhibit cycle-averaged drag reduction, but net energy savings are not achieved in every case. The energetic benefit depends on the duration and timing of low drag relative to the bulk flow rate. Within each Reynolds-number range, the plateau case extends the low-drag interval and improves both measures.

\begin{bmhead}[Acknowledgements.]
The authors used ChatGPT and Codex (OpenAI) for Chinese-to-English translation,
language editing, collation of previously recorded numerical values, and generation
of post-processing scripts. Details of the AI-assisted post-processing and its
verification are provided in \S\ref{sec:postprocessing}.
The authors take full responsibility for the final manuscript.
\end{bmhead}

\begin{bmhead}[Funding.]
This work was supported by the National Natural Science Foundation of China
(grant numbers 12372270 and U2341242).
\end{bmhead}

\begin{bmhead}[Declaration of interests.]
The authors report no conflict of interest.
\end{bmhead}

\begin{bmhead}[Supplementary material.]
Supplementary movies~1--3 are available with the online version of this article.
\end{bmhead}

\bibliographystyle{jfm}
\bibliography{references}

\end{document}